\documentclass{aa}  

\usepackage{graphicx}
\usepackage{txfonts}

\usepackage[breaklinks,colorlinks,citecolor=blue]{hyperref}
\usepackage[all]{hypcap}
\usepackage{ulem}

\usepackage{lmodern}

\bibpunct{(}{)}{;}{a}{}{,} 

\newif\iffigs
\newif\iffigscl
\newif\iffigstest
\newif\iflabs

\figstrue

\figscltrue

\figstestfalse

\labsfalse

\newcommand{\GamSig}{$\Gamma$-$\Sigma_{\rm SFR}\;$}
\newcommand{\GamI}{$\Gamma$-$1\;$}

\begin{document} 


\newif\iffigs

\figstrue

\newcommand{\itii}[1]{{#1}}
\newcommand{\franta}[1]{\textbf{\color{green} #1}}
\newcommand{\frantaii}[1]{\textbf{\color{yellow} #1}}
\newcommand{\itiitext}[1]{{#1}}

\newcommand{\eq}[1]{eq. (\ref{#1})}
\newcommand{\eqp}[1]{(eq. \ref{#1})}
\newcommand{\eqq}[1]{eq. \ref{#1}}
\newcommand{\eqb}[2]{eq. (\ref{#1}) and eq. (\ref{#2})}
\newcommand{\eqc}[3]{eq. (\ref{#1}), eq. (\ref{#2}) and eq. (\ref{#3})}
\newcommand{\refs}[1]{Sect. \ref{#1}}
\newcommand{\reff}[1]{Fig. \ref{#1}}
\newcommand{\reft}[1]{Table \ref{#1}}

\newcommand{\datum} [1] { \noindent \\#1: \\}
\newcommand{\pol}[1]{\vspace{2mm} \noindent \\ \textbf{#1} \\}
\newcommand{\code}[1]{\texttt{#1}}
\newcommand{\figpan}[1]{{\sc {#1}}}

\newcommand{\nbdvi}{\textsc{nbody6} }
\newcommand{\nbdvid}{\textsc{nbody6}}
\newcommand{\flash}{\textsc{flash} }
\newcommand{\flashd}{\textsc{flash}}
\newcommand{\sfe}{\mathrm{SFE}}
\newcommand{\sfei}[1]{$\mathrm{SFE} = #1$ \%}
\newcommand{\sfr}{\mathrm{SFR}}
\newcommand{\mum}{$\; \mu \mathrm{m} \;$}
\newcommand{\rop}{$\rho$ Oph }
\newcommand{\HT}{$\mathrm{H}_2$}
\newcommand{\Halpha}{$\mathrm{H}\alpha \;$}
\newcommand{\HI}{H {\sc i} }
\newcommand{\HII}{H {\sc ii} }
\renewcommand{\deg}{$^\circ$}

\newcommand{\dd}{\mathrm{d}}
\newcommand{\acosh}{\mathrm{acosh}}
\newcommand{\sign}{\mathrm{sign}}
\newcommand{\cex}{\mathbf{e}_{x}}
\newcommand{\cey}{\mathbf{e}_{y}}
\newcommand{\cez}{\mathbf{e}_{z}}
\newcommand{\cer}{\mathbf{e}_{r}}
\newcommand{\ceR}{\mathbf{e}_{R}}

\newcommand{\llg}[1]{\log_{10}#1}
\newcommand{\pder}[2]{\frac{\partial #1}{\partial #2}}
\newcommand{\pderrow}[2]{\partial #1/\partial #2}
\newcommand{\nder}[2]{\frac{\dd #1}{\dd #2}}
\newcommand{\nderrow}[2]{{\dd #1}/{\dd #2}}

\newcommand{\Cmiii}{\, \mathrm{cm}^{-3}}
\newcommand{\Gcmii}{\, \mathrm{g} \, \, \mathrm{cm}^{-2}}
\newcommand{\Gcmiii}{\, \mathrm{g} \, \, \mathrm{cm}^{-3}}
\newcommand{\Kms}{\, \mathrm{km} \, \, \mathrm{s}^{-1}}
\newcommand{\Si}{\, \mathrm{s}^{-1}}
\newcommand{\Esi}{\, \mathrm{erg} \, \, \mathrm{s}^{-1}}
\newcommand{\Ee}{\, \mathrm{erg}}
\newcommand{\Yr}{\, \mathrm{yr}}
\newcommand{\Kyr}{\, \mathrm{kyr}}
\newcommand{\Myr}{\, \mathrm{Myr}}
\newcommand{\Gyr}{\, \mathrm{Gyr}}
\newcommand{\Msun}{\, \mathrm{M}_{\odot}}
\newcommand{\Rsun}{\, \mathrm{R}_{\odot}}
\newcommand{\Pc}{\, \mathrm{pc}}
\newcommand{\Kpc}{\, \mathrm{kpc}}
\newcommand{\Mpc}{\, \mathrm{Mpc}}
\newcommand{\Sd}{\Msun \, \Pc^{-2}}
\newcommand{\Ev}{\, \mathrm{eV}}
\newcommand{\Kk}{\, \mathrm{K}}
\newcommand{\Au}{\, \mathrm{AU}}
\newcommand{\Mas}{\, \mu \mathrm{as}}
\newcommand{\Mspy}{\, \mathrm{M}_{\odot}\, \, \Yr^{-1}}

   \title{Do the majority of stars form as gravitationally unbound?}

   \authorrunning{Dinnbier, Kroupa and Anderson}
   \titlerunning{The fraction of stars forming in bound star clusters}

   \author{Franti\v{s}ek Dinnbier\inst{1},
          Pavel Kroupa\inst{1,2}
          \and
          Richard I. Anderson\inst{3}
          }

   \institute{Astronomical Institute, Faculty of Mathematics and Physics, Charles University, V Hole\v{s}ovi\v{c}k\'{a}ch 2, 180 00 Praha 8, Czech Republic \\
             \email{dinnbier@sirrah.troja.mff.cuni.cz}
         \and
             Helmholtz-Institut f\"{u}r Strahlen- und Kernphysik, University of Bonn, Nussallee 14-16, 53115 Bonn, Germany \\
             \email{pavel@astro.uni-bonn.de}
         \and
             Institute of Physics, Laboratory of Astrophysics, \'Ecole Polytechnique F\'ed\'erale de Lausanne (EPFL), Observatoire de Sauverny, 1290 Versoix, Switzerland \\
             \email{richard.anderson@epfl.ch}
             }

   \date{Received \today; accepted ??}

  \abstract
  {Some of the youngest stars (age $\lesssim 10 \Myr$) are clustered, while many others are observed scattered throughout star forming regions or in complete 
   isolation. It has been intensively debated whether the scattered or isolated stars originate in star clusters, or if they form truly isolated, which could help 
   constrain the possibilities how massive stars are formed.
  }
  {We adopt the assumption that all stars form in gravitationally bound star clusters embedded in molecular cloud cores (\GamI model), 
   which expel their natal gas early after their formation, 
   and compare the fraction of stars found in clusters with observational data. 
  }
  {
   The star clusters are modelled by the code \nbdvid, which includes binary stars, stellar and circumbinary evolution, gas expulsion, and the 
   external gravitational field of their host galaxy.
  }
  {
   We find that small changes in the assumptions in the current theoretical model estimating the fraction, $\Gamma$, of stars forming 
   in embedded clusters have a large influence on the results, and we present a counterexample as an illustration. 
   This calls into question theoretical arguments about $\Gamma$ in embedded clusters, 
   and it suggests that there is no firm theoretical ground for low $\Gamma$ in galaxies with lower star formation rates (SFRs).
   Instead, the assumption that all stars form in embedded clusters is in agreement with observational data for the youngest stars (age $\lesssim 10 \Myr$). 
   In the \GamI scenario, the observed fraction of the youngest stars in clusters increases with the SFR only weakly; 
   the increase is caused by the presence of more massive clusters in galaxies with higher SFRs, which release fewer stars to the field in proportion to their mass.
   The \GamI model yields a higher fraction of stars in clusters for older stars (age between $10 \Myr$ and $300 \Myr$) than what is observed. 
   This discrepancy can be caused by initially less compact clusters and/or a slightly lower star formation efficiency than originally assumed in the \GamI model, 
   or by interactions of the post gas expulsion revirialised open clusters with molecular clouds.
  }
  {}


   \maketitle
%

\section{Introduction}


Star formation takes place in the densest parts of molecular clouds. 
The youngest stars are usually found to be concentrated in embedded star clusters \citep{Lada1991,Lada2003,Megeath2016} or more dispersed 
throughout OB associations \citep{Blaauw1964}. 
This opens two broad scenarios for the conditions for star formation. 

In the first scenario, all stars form in gravitationally bound embedded 
clusters \citep[e.g.][]{Lada1984,Kroupa1995b,Lada2003,Porras2003,Kroupa2001b,Banerjee2017,Gonzalez2020}, 
which form throughout a molecular cloud, which as a whole, might not be necessarily gravitationally bound, 
but merely a condensation in the interstellar medium. 
The clusters expand and lose a substantial fraction of their stars as the result of the expulsion of the non-star forming gas. 
Cores of some of these clusters survive the violent gas expulsion event \citep{Lada1984,Boily2003b} and they are observed as gas-free open star clusters, 
while the escaping stars are receding out of clusters, and they are observed as unbound OB associations. 
After gas expulsion, the surviving cluster still evaporates and ejects stars, albeit at a slower rate than during gas expulsion. 
Dynamical ejections are particularly important for the most massive stars in the cluster, where the process is 
able to kick tens of percent of OB stars from the cluster \citep{Fujii2011,Oh2015,Oh2016,Wang2019}. 

In the second scenario, stars form throughout the molecular cloud in hierarchical structures following the density field created by interstellar turbulence 
\citep[e.g.][]{Preibisch1999,Elmegreen2000,Clark2005,Elmegreen2006}. 
While some star formation occurs in initially bound clusters, a substantial amount of star formation occurs in initially non gravitationally bound groups. 
Thus, unlike in the first scenario, OB associations are gravitationally unbound from their formation. 

In this context it is useful to define the fraction of star formation $\Gamma = \rm{CFR}/\rm{SFR}$ occurring in initially gravitationally bound 
embedded clusters. 
The CFR is the star formation rate within all bound clusters in the galaxy, and the SFR is the total star formation rate of the galaxy.

The two scenarios have deep implications not only for the formation of bound star clusters, but also for massive star formation, 
which still lacks a full understanding \citep{Zinnecker2007}. 
In the first scenario, massive stars are clustered, and they compete for the available gas with each other \citep{Bonnell1997,Bonnell2001}, 
lose their specific angular momentum in the process, which naturally leads to their primordial mass segregation. 
In clusters, more massive stars can be also produced by their mergers still in the pre-main-sequence stage \citep{Bonnell1998}. 
In contrast, the second scenario is based on forming massive stars in isolation, a process for which there 
is not much observational evidence \citep{Selier2011}.

In the Galaxy, observations suggest that the majority ($\Gamma = 0.5$ to $1.0$) of star formation occurs 
in embedded clusters \citep[e.g.][]{Carpenter1995,Carpenter2000,Lada2003,Porras2003,Bressert2010,Winston2020}.
An independent piece of evidence for $\Gamma$ close to $1$ comes from the observed lower binary fraction of field stars in contrast to the 
binary fraction in star forming regions, which can be explained 
as a result of dynamical processing within clusters \citep{Kroupa1995a,Kroupa1995b,Marks2011}.
When focusing on O stars, only $20$ \% to $50$\% of them are located in the field (i.e. outside clusters or OB associations) \citep{Mason1998}, but 
many of the field O stars are unambiguously identified as runaway stars originating from bound systems \citep[e.g.][]{Blaauw1964,Stone1991,Tetzlaff2011,Gvaramadze2012}, 
further decreasing the upper limit of the truly isolated massive star formation. 

In contrast, some observations of external galaxies point to a substantially lower values of $\Gamma$ 
of the order of $0.1$ \citep[e.g.][]{Goddard2010,Adamo2011,Johnson2016}.
\citet{Kruijssen2012} suggests a semi-analytic model, according to which $\Gamma$ increases with the star formation rate of the galaxy per 
unit area $\Sigma_{\rm SFR}$. 
However, later \citet{Chandar2017} found substantially larger values of $\Gamma$ in galaxies with low $\Sigma_{\rm SFR}$, and also no clear 
dependence of $\Gamma$ on the $\Sigma_{\rm SFR}$.
They suggest that the dependency previously reported was caused by an age biased sample. 
Likewise, \citet{Fensch2019} report large values of $\Gamma$ in tidal dwarf galaxies located around galaxy NGC~5291, which differ by $\approx 3 \sigma$ from 
the estimate by \citet{Kruijssen2012}.
\citet{Stephens2017} select a sample of well-isolated massive young stellar objects in the Large Magellanic Cloud (LMC), 
previously thought to be massive stars formed in isolation, and find that each of them is 
surrounded by a star cluster of $\gtrsim 100$ stars, indicating that the vast majority ($\gtrsim 95$ \%) of massive stars form in clusters instead of in isolation. 
Thus, it is possible that the vast majority (if not all) massive stars found currently in isolation in external galaxies, are in reality 
runaway stars, or they are surrounded by a small star cluster consisting of lower mass stars, which is not resolvable within the given survey.

In the present paper, we assume that all stars form in gravitationally bound embedded clusters ($\Gamma = 1$), which evolve due to their cluster dynamics, 
gas expulsion, stellar evolution from a realistic initial stellar mass function (IMF), and the tidal field of their host galaxy. 
We study the fraction of stars to be found within the clusters as a function of the cluster age, mass, orbital radius within the galaxy, 
and we synthesize these results for a population of star clusters for a galaxy of a given global star formation rate in the framework 
of the integrated galactic initial mass function (IGIMF), and provide a comparison to observational data.
Our assumption of $\Gamma = 1$ contrasts with the theoretical value of $\approx 0.1$ as expected for a Milky Way-like galaxy from the 
theoretical model of \citet{Kruijssen2012}, which provides $\Gamma$ as a function of the properties of the galaxy, and which 
favours isolated massive star formation in galaxies with lower star formation rates.
We question some of the assumptions used in the theoretical model, 
and we illustrate on a simple example its main limitations.

\section{The current model for $\Gamma$ and its limitations}

\label{sLimitations}




\subsection{Overview of the model}

\label{ssGamSigOverview}

\citet{Kruijssen2012} (hereafter K12) presented the first (and the only known to us) theoretical framework to infer 
$\Gamma$ from the properties of the interstellar 
medium (ISM) of the particular galaxy. 
The model assumes that the probability density function $\dd p/\dd x$ of the ISM mass density is log-normal, and that star formation proceeds at a constant rate per free-fall time, 
so the denser gas produces stars more rapidly.
The variable $x$ denotes the normalised local density contrast, $x = \rho/\rho_{\rm ISM}$, where $\rho_{\rm ISM}$ is the midplane density. 
Star formation is terminated either by gas exhaustion, by supernova feedback, or it still continues up to the age of interest.
If star formation terminates by gas exhaustion, the final star formation efficiency (SFE $\equiv$ stellar mass/(stellar plus gaseous mass)) 
is set to $0.5$, which is motivated by the reduction of the stellar mass by protostellar outflows.
If star formation is terminated by feedback from newly formed stars, the termination time is calculated by 
balancing the thermal pressure due to feedback with the turbulent pressure of the ISM. 

The fraction of stars $\gamma$, which form in gravitationally bound clusters is estimated to be linearly proportional to the SFE, i.e. $\gamma = 2\, \sfe$ 
(the factor $2$ is to reach $\gamma = 1$ for the maximum allowed $\sfe$ of $0.5$). 
The clusters are subjected to tidal fields of passing interstellar clouds, which dissolve all clusters, which formed at normalised density below $x_{\rm cce}$ 
on a time-scale shorter than the specified age limit, which is chosen to be $10 \Myr$; clusters formed in gas above this density are assumed to be intact. 
Then, the fraction of stars which form in gravitationally bound clusters is calculated as 
\begin{equation}
\Gamma = \frac{\int_{x_{\rm cce}}^{\infty} \gamma(x) \sfe(x) x (\dd p/\dd x) \dd x}{\int_{-\infty}^{\infty} \sfe(x) x (\dd p/\dd x) \dd x}. 
\label{eGammaDef}
\end{equation}

Since all the dependent variables under the integral sign as well as $x_{\rm cce}$ depend implicitly on the galactic parameters 
$\Sigma_{\rm g}$ (mean gas surface density), $\Omega$ (angular frequency) and $Q$ (Toomre parameter), 
$\Gamma$ can be expressed as a function of these three galactic parameters. 
Further simplification is obtained by expressing the star formation rate $\Sigma_{\rm SFR}$ of the galaxy 
by $\Sigma_{\rm g}$ according to the Schmidt star formation law \citep{Schmidt1959,Kennicutt1998}. 
Taking typical values for the two other parameters ($\Omega$ and $Q$), 
K12 arrived at a formula for $\Gamma$ as a function of $\Sigma_{\rm SFR}$ only. 
Because of this property, we refer to this model as the \GamSig model hereafter in this work.

\subsection{Limitations of the model}

\label{ssGamSigLimitations}


The \GamSig model provides a valuable theoretical formulation for understanding the functional dependence of physical quantities.
However, its derivation requires several simplifications or assumptions that require further consideration because they significantly affect model predictions. In particular:

\begin{itemize}
\item
Probably the most significant simplification is the absence of the threshold for star formation of $\approx 120 \Msun\, \, \Pc^{-2}$
\citep{Lada2009,Lada2010,Heiderman2010}, which suggests that star formation in the Galaxy
is absent at gas densities below $n (\mathrm{H}_2) \approx 10^4 \Cmiii$ ($\approx 300 \Msun \, \Pc^{-3}$).
For the normalised density contrast, this means that star formation occurs only at $x_{\rm lim} \gtrsim 10^4$ (for a Milky Way-like galaxy),
while the \GamSig model provides a much lower threshold of $x_{\rm cce} \approx 10^2$ for bound cluster formation and survival (Sect. 2.7.3 in K12).
The model even considers scattered star formation at gas densities below $x = 10^2$ (cf. their figure 1), i.e. at densities $n (\mathrm{H}_2) \lesssim 10^2 \Cmiii$.
The low star formation threshold results in many stars forming as gravitationally unbound, leading to low values of $\Gamma$.
\item
The \GamSig model assumes (Sect. 2.5.2 in K12) that star formation proceeds at a constant rate until it is quenched by supernovae (which occur at earliest 
at $3 \Myr$ after massive star formation started), or alternatively by radiation pressure. 
This neglects the influence of important early forms of stellar feedback, e.g. the photoionising radiation and stellar winds, which have a strong 
impact on the surrounding gas, substantially dispersing the natal cloud before the first supernova occurs \citep[e.g.][]{Rogers2013,Gavagnin2017,Haid2018,Dinnbier2020c}. 
Neglecting early stellar feedback likely overestimates the value of $\sfe(x)$.
\item
The functional form of the probability density function of self-gravitating star forming gas, which is a power-law function of a slope close
to $-2$ \citep[e.g.][]{Kritsuk2011,Schneider2013,Lombardi2015,Chen2018} is very different from the log-normal distribution of non-self gravitating
turbulent ISM \citep{Vazquez-Semadeni1994,Padoan1997,Kritsuk2011} as assumed in the \GamSig model.
\item
As the collapse of the star forming cloud proceeds, its free-fall time-scale decreases because  density increases. 
Conversely, the \GamSig model assumes a constant SFR with time (Eq. 22 in K12).
\item
To derive the density threshold $x_{\rm cce}$ (Sect. 2.7 in K12), it is assumed that the ISM velocity dispersion $\sigma_{\rm g}$ is constant regardless of the distance between the clumps.
This is a reasonable approximation for the relative velocity between two molecular clouds. However, for the scale-free ISM within a given cloud, $\sigma_{\rm g}$ depends on distance $l$ between the
target cluster and the clump according to the Larson relations \citep[e.g.][]{Larson1981,Heyer2009}.
\item The \GamSig model assumes that all molecular clouds are of the same mass (eq. 35 in K12).
\item
The \GamSig model does not consider star cluster dynamics, which can eject a substantial fraction of 
young massive stars from clusters even during the first several Myr of 
the cluster's existence \citep{Perets2012,Oh2015,Wang2019}, thus underestimating $\Gamma$. 
\item
Expulsion of the non-star forming gas is not considered. 
Instead, it is assumed that most of the gas is consumed by star formation, which is not reproduced 
by many of the more recent star forming simulations \citep[e.g.][]{Colin2013,Gavagnin2017,Gonzalez2020}.
\end{itemize}

In order to illustrate the strong dependence of the \GamSig model on its assumptions, we consider a very simple model for star formation that agrees with most observational facts, yet leads to completely different results than the \GamSig model.
We stress that we use this as an illustration only and do not claim that star formation occurs according to this toy model. Rather, we argue that a
radically different approach is needed to tackle the highly non-linear process of clustered star formation from a theoretical perspective.
Consider that all stars form only at density $x > x_{\rm lim}$, and star formation occurs in gravitationally bound clusters if $x > x_{\rm cce}$, and
in unbound associations if $x < x_{\rm cce}$.
It means that $\gamma(x) = 1$ for $x > x_{\rm cce}$ and $0$ otherwise.
We further assume that $\sfe(x) = 1/3$ regardless of $x$ \citep[e.g.][]{Geyer2001,Kroupa2001b,Megeath2016,Banerjee2017,Geen2018}.
From \eq{eGammaDef}, these assumptions lead to $\Gamma = 1$ for galaxies with $x_{\rm cce} < x_{\rm lim}$, and $\Gamma$ decreasing 
with $x_{\rm cce}$ for $x_{\rm cce} > x_{\rm lim}$, with $x_{\rm cce} = x_{\rm lim}$ at $\rm{log}_{10} (\rho_{\rm ISM} $[$\Gcmiii$]$) \approx -21.8$
(taking aside the weaker dependence on $\Omega$ and $Q$, we note that $x_{\rm cce}$ can be used as a proxy for $\Sigma_{\rm SFR}$).
This model is in a stark contrast to the \GamSig model, where $\Gamma \approx 0.1$ for galaxies with low $x_{\rm cce}$ of $\approx 10^2$,
which increases with $x_{\rm cce}$ towards $\approx 0.7$.

By choosing different functional forms for $\gamma(x)$ and $\sfe(x)$, neither of which are well constrained, it is easy to find an almost arbitrary dependence  
of $\Gamma$ on $x_{\rm cce}$, and therefore on $\Sigma_{\rm SFR}$. 
The extreme sensitivity on initial assumptions
\footnote{The sensitivity to the choice of the particular functional forms was mentioned by \citet[their appendix D]{Kruijssen2012}, 
although its influence on results was not investigated or discussed in detail.}
%
is caused by recursively substituting 
equations into another, several of which are non-linear or contain terms of uncertain importance.

This behaviour is typical for chaotic systems, which cannot be understood by linearising or simplifying relevant equations. Thus,
redoing the analysis for the \GamSig model in the same way will not yield more reliable results even if the aforementioned issues are resolved.

The above toy model illustrates that even minor changes in the assumptions of the \GamSig model lead to drastically different outcomes 
and thus questions the previously suggested theoretical result that most stars form in gravitationally unbound entities in a Milky Way-like galaxy.

In the following, we have therefore adopted the opposite approach, namely that all stars form in gravitationally bound embedded clusters (i.e. $\Gamma = 1$), 
to investigate the currently observed fraction of bound stars by dynamical N-body simulations.
The assumption of $\Gamma = 1$ is motivated by observational surveys of pre-main sequence stars 
in molecular clouds \citep[e.g.][]{Lada1991,Megeath2016,Joncour2018}, 
the distribution functions of binary stars in the Galactic field in comparison to star forming regions \citep[e.g.][]{Reipurth1993,Leinert1993,Kroupa1995a}, 
and the general argument that a star forming molecular cloud core is observed to always contain significantly more gas than the weight of 
a single late-type star \citep{Alves2007,Goodwin2008}. 
We refer to the present model as \GamI model to distinguish it from the \GamSig model.

Further assumptions of the \GamI model are that the clusters are initially compact (half-mass radii $r_{\rm h}$ of $\lesssim 1 \Pc$ 
for clusters of mass $M_{\rm ecl} \lesssim 10^4 \Msun$).
The initial cluster mass function is a power-law \eqp{eicmf}, whose slope and upper mass limit 
follows from the galactic star forming rate as obtained from the IGIMF theory \citep{Kroupa2003b}. 
The clusters form with $\sfe = 1/3$, and they expel their non-star forming gas on the timescale of $r_{\rm h}/10 \Kms \approx 0.05 \Myr$
at the cluster age of $0.6 \Myr$. 

The particular values of the parameters in the \GamI model might be adjusted in the future when more 
accurate observational data become available. 
This iteration could provide another constraint on the birth properties of star clusters.

\section{Model description and numerical method}

\label{sModelNum}

\subsection{Model description}

\label{ssModel}

The main results of this article are based on the models presented in \refs{sssStandardModels} and correspond 
to the most probable initial conditions for embedded star clusters (standard models).
To assess the dependence on the particular choice of initial assumptions, we compute additional models to investigate the influence of the 
initial cluster radius, the SFE, and primordial mass segregation (\refs{sssAdditionalModels}).

\subsubsection{Standard models}

\label{sssStandardModels}

Observed open clusters are well described by the Plummer model \citep[e.g.][]{Roser2011,Roser2019} and star forming molecular cloud
filaments have Plummer-like cross sections \citep[e.g.][]{Andre2014}, which motivates us to use this description as initial conditions for present models.  
The Plummer model is described by its stellar mass, $M_{\rm ecl}$, and Plummer parameter, $a_{\rm ecl}$.
The Plummer parameter is related to the cluster mass by the mass-radius relation of \citet{Marks2012}, where 
\begin{equation}
\frac{a_{\rm ecl}}{\Pc} = 0.077 \left( \frac{M_{\rm ecl}}{\Msun} \right)^{1/8}.
\label{eAMeclStandard}
\end{equation}
The stellar masses of the clusters in our model grid are $50 \Msun$, $100 \Msun$, \ldots $6400 \Msun$, which approximately covers the mass range of observed star clusters in 
the Galaxy and the galaxy M~31. 
To obtain better statistics, we simulate the $50 \Msun$ clusters $256$ times with different random seed, $100 \Msun$ clusters $128$ times etc., so the 
$6400 \Msun$ clusters are simulated twice. 

Stellar masses are sampled from the \citet{Kroupa2001a} model of the initial mass function (IMF) in the mass range $(0.08 \Msun, m_{\rm max})$. 
To allow only the more massive clusters to form massive stars, we adopt the $m_{\rm max} - M_{\rm ecl}$ relation from \citet{Weidner2010} 
(see also \citealt{Elmegreen1983,Elmegreen2000} for earlier works on the same topic). 
Initial conditions for the clusters are generated by the software package \textsc{mcluster}, which is described in \citet{Kupper2011}.
Initially, all stars are in binaries. 
Stars with mass below $3 \Msun$ have the distribution of orbital periods according to \citet{Kroupa1995a}, while 
stars more massive than that have the distribution of orbital periods of \citet{Sana2012}, which generally results in more compact systems. 

The expulsion of the gas which was not consumed by the forming stars has a profound impact on cluster dynamics 
(e.g. \citealt{Tutukov1978,Lada1984,Goodwin1997,Kroupa2001b,Geyer2001,Banerjee2013,Banerjee2017}). 
We approximate the gaseous component by a Plummer potential with $a_{\rm ecl}$ identical to that of the stellar component, 
and of mass $2 M_{\rm ecl}$, so we assume the SFE to be $33$ \%. 
When the embedded phase ends at time $t_{\rm d} = 0.6 \Myr$, the mass of the gaseous component is reduced exponentially 
on the time scale $\tau_{\rm M} = 1.3 a_{\rm ecl}/10 \Kms \approx 0.04 \Myr$; it is the same approximation as used by \citet{Kroupa2001b}. 

To study stellar escape from star clusters realistically, one needs to include the tidal forces due to the host galaxy. 
Accordingly, we adopt the galactic potential of \citet{Allen1991}, which consists of the disc, halo and the central part of the galaxy. 
The modelled clusters orbit the galaxy on circular trajectories of radii $R_{\rm g} = 4 \Kpc$, $R_{\rm g} = 8 \Kpc$ and $R_{\rm g} = 12 \Kpc$. 
All these models are calculated for stars of Solar metallicity ($Z=0.014$). 
In addition, we calculate clusters with the subsolar metallicities of $Z=0.006$ and $Z=0.002$ orbiting at $R_{\rm g} = 8 \Kpc$.

\subsubsection{Models for exploring the sensitivity on initial conditions}

\label{sssAdditionalModels}

We study the influence of the particular functional form between the Plummer parameter and cluster mass, i.e. $a_{\rm ecl} = a_{\rm ecl}(M_{\rm ecl})$. 
In these models, we assume that
\begin{equation}
\frac{a_{\rm ecl}}{\Pc} = 0.034 \left( \frac{M_{\rm ecl}}{\Msun} \right)^{1/3},
\label{eAMecl}
\end{equation}
which presents a stronger mass dependence than in the standard models \eqp{eAMeclStandard}. 
Normalisation of \eq{eAMecl} is chosen so that the least massive clusters ($M_{\rm ecl} = 50 \Msun$) have the same half-mass radius ($r_{\rm h} = 0.163 \Pc$) 
as clusters of the same mass in standard models. 
The most massive clusters in our models ($M_{\rm ecl} = 6400 \Msun$) have a half-mass radius of $0.79 \Pc$ according to \eq{eAMecl}, 
while it is $0.30 \Pc$ in standard models. 
The form of \eq{eAMecl} means that embedded clusters have the same density regardless of their mass. 
In discussion, we refer to these models as "A" models, while standard models are referred to "S" models.

Another quantity of interest is the SFE, which we decrease to $0.25$. 
Otherwise, the clusters are identical as in the standard models. 
In discussion, we refer to these models as "F" models.
"A" and "F" models are calculated for the same library of clusters as the standard models with $R_{\rm g} = 8\Kpc$ and $Z = 0.014$, but 
we realised all clusters of mass $\leq 400 \Msun$ only $32$ times.

We also study the influence of primordial mass segregation (\refs{ssInfluenceICs}).
These models are generated according to the recipe of \citet{Subr2008} with mass segregation index $S = 0.5$.
This results in highly mass segregated clusters; for example, the half-mass radius of M stars is $0.4 \Pc$, while the half-mass radius of B stars is $0.1 \Pc$.
Otherwise, the models are identical to the standard models.  
The role of primordial mass segregation is studied only in $M_{\rm ecl} = 400 \Msun$ clusters orbiting at $R_{\rm g} = 8 \Kpc$ and with $Z = 0.014$. 
Similar to the standard models, we simulate each $400 \Msun$ cluster $32$ times with different random seeds.
These models are labelled "M" models.


\subsection{Numerical method}

\label{ssNumerics}

The clusters are integrated by the code \nbdvid, which uses a 4th order Hermite predictor-corrector \citep{Makino1991}, quantised time-steps with 
the Ahmad-Cohen method \citep{Ahmad1973,Makino1992}. 
Compact subsystems of two, three and more stars are handled by regularising techniques \citep{Kustaanheimo1965,Aarseth1974a,Mikkola1990}.
Metallicity-dependent stellar evolution is included \citep{Tout1996,Hurley2000} as well as an approximation to circumbinary evolution including 
Roche mass transfer and common-envelope evolution \citep{Hurley2002}.
A detailed description of the algorithms used in \nbdvi is provided in \citet{Aarseth2003}.

\section{Results}

\label{sResults}

For the purpose of this work, we define a star cluster as a group of at least $10$ centre of mass bodies which are gravitationally bound. 
The fraction $f_{\rm IC}$ of stars in clusters is the number of stars in clusters $N_{\rm in}$ divided 
by the sum of stars in clusters $N_{\rm in}$ and in the field $N_{\rm out}$.
We discuss $f_{\rm IC}$ in different contexts: $f_{\rm IC} (M_{\rm ecl})$, $f_{\rm IC, pop}$ and $f_{\rm IC} (m)$
is the fraction of stars in clusters for clusters of mass $M_{\rm ecl}$, for the whole population of star clusters which formed according to a given 
initial cluster mass function, and for stars of stellar mass $m$ formed by a whole population of star clusters, respectively. 
The time averaged fraction of stars in age interval $\Delta t$ formed by the whole cluster population is denoted $\overline{f}_{\rm IC,pop} (\Delta t)$, 
and the likely observed fraction of these stars in clusters is denoted $\overline{f}^{\rm obs}_{\rm IC,pop} (\Delta t)$.

We assume two idealised scenarios for the star formation history. 
In the first one, all star clusters are formed in a single starburst of a short ($\lesssim 10 \Myr$) duration with no star formation afterwards 
(Sect. \ref{ssGalaxyWide} and \ref{ssIGIMF}). 
In the second one, star formation proceeds continuously with a constant star formation rate (SFR; \refs{ssContSFR}).
A comparison with observational data and discussion of the most likely value of $\Gamma$ is discussed in \refs{ssCombObs}.

\subsection{The dependence on cluster mass}

\label{ssClusterMass}

\iffigscl
\begin{figure}
\includegraphics[width=\columnwidth]{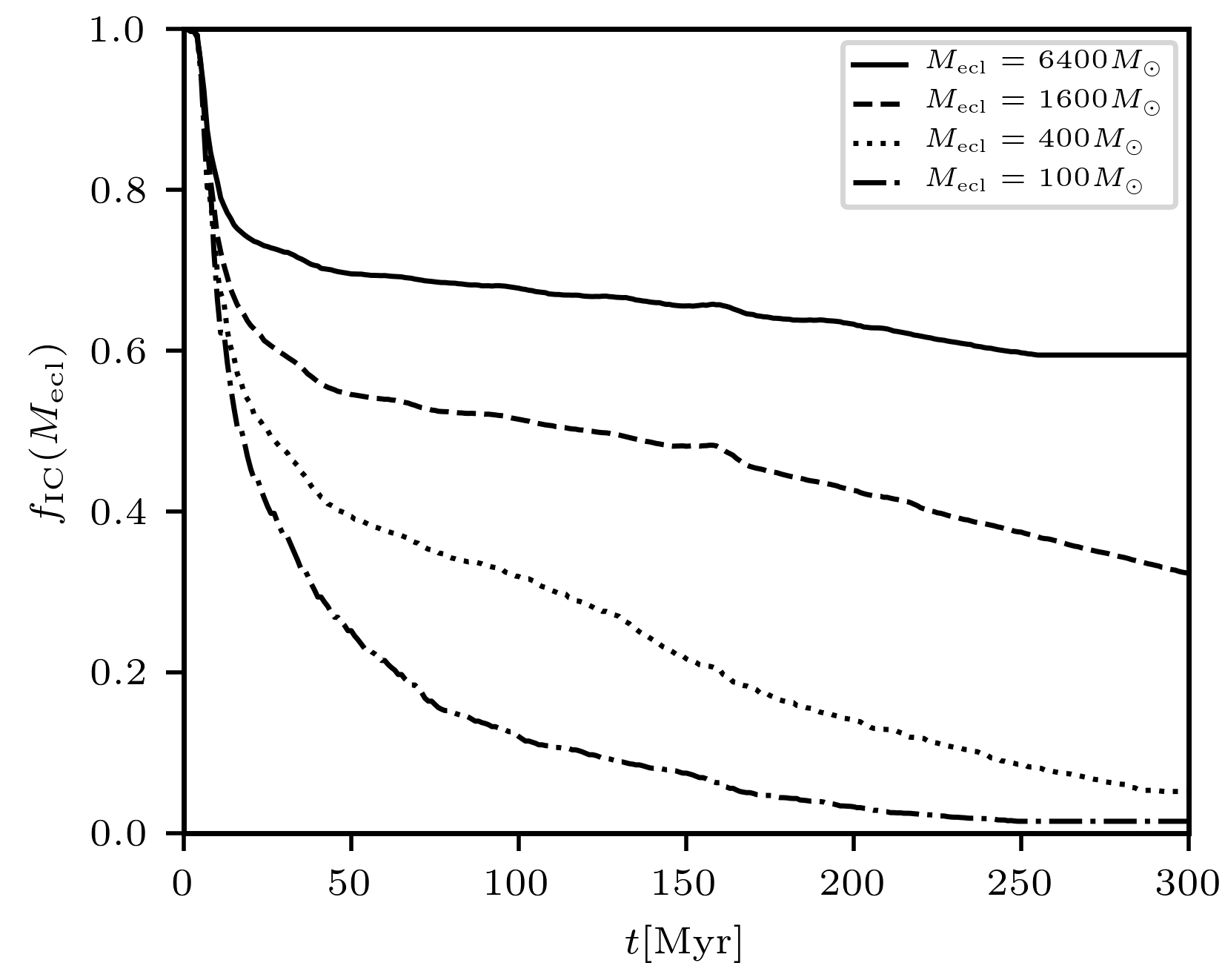}
\caption{The fraction of all stars which are present in their birth clusters at a given age $t$ 
for clusters of initial mass $M_{\rm ecl} = 100 \Msun$ (dash-dotted line), 
$M_{\rm ecl} = 400 \Msun$ (dotted line), $M_{\rm ecl} = 1600 \Msun$ (dashed line), and $M_{\rm ecl} = 6400 \Msun$ (solid line).
}
\label{fIndividualClStars}
\end{figure} \else \fi

The fraction $f_{\rm IC} (M_{\rm ecl})$ of stars in clusters for clusters of different mass, $M_{\rm ecl}$, 
as a function of cluster age, $t$, is shown in \reff{fIndividualClStars}. 
These clusters orbit the galaxy at the radius $R_{\rm g} = 8 \Kpc$, and have a metallicity $Z = 0.014$. 
This plot includes stars of all spectral types. 
The rapid decrease of $f_{\rm IC} (M_{\rm ecl})$ during the first $\approx 15 \Myr$ is caused by the reaction of the cluster to gas expulsion. 
After that, the decrease of $f_{\rm IC} (M_{\rm ecl})$ slows down, where more massive clusters retain more stars at a given time. 
By $300 \Myr$, all the clusters with initial mass $M_{\rm ecl} \lesssim 400 \Msun$ released more than $95$\% of their stars to the field, while 
clusters with $M_{\rm ecl} = 6400 \Msun$ released only $40$\% of their stars into the field. 
An estimate of the fraction of stars in clusters, which is likely to be observed, is shown in the left panel of \reff{fobserved}.

The decrease of $f_{\rm IC} (M_{\rm ecl})$ with decreasing $M_{\rm ecl}$ is expected as lower mass clusters are impacted more by gas expulsion 
\citep{Baumgardt2007}, and have a shorter median two-body relaxation time-scale which leads 
to more rapid evaporation and cluster dissolution \citep{Baumgardt2003,Lamers2005}.

\subsection{The galaxy wide fraction of stars in clusters in a single starburst} 

\label{ssGalaxyWide}

\iffigscl
\begin{figure}
\includegraphics[width=\columnwidth]{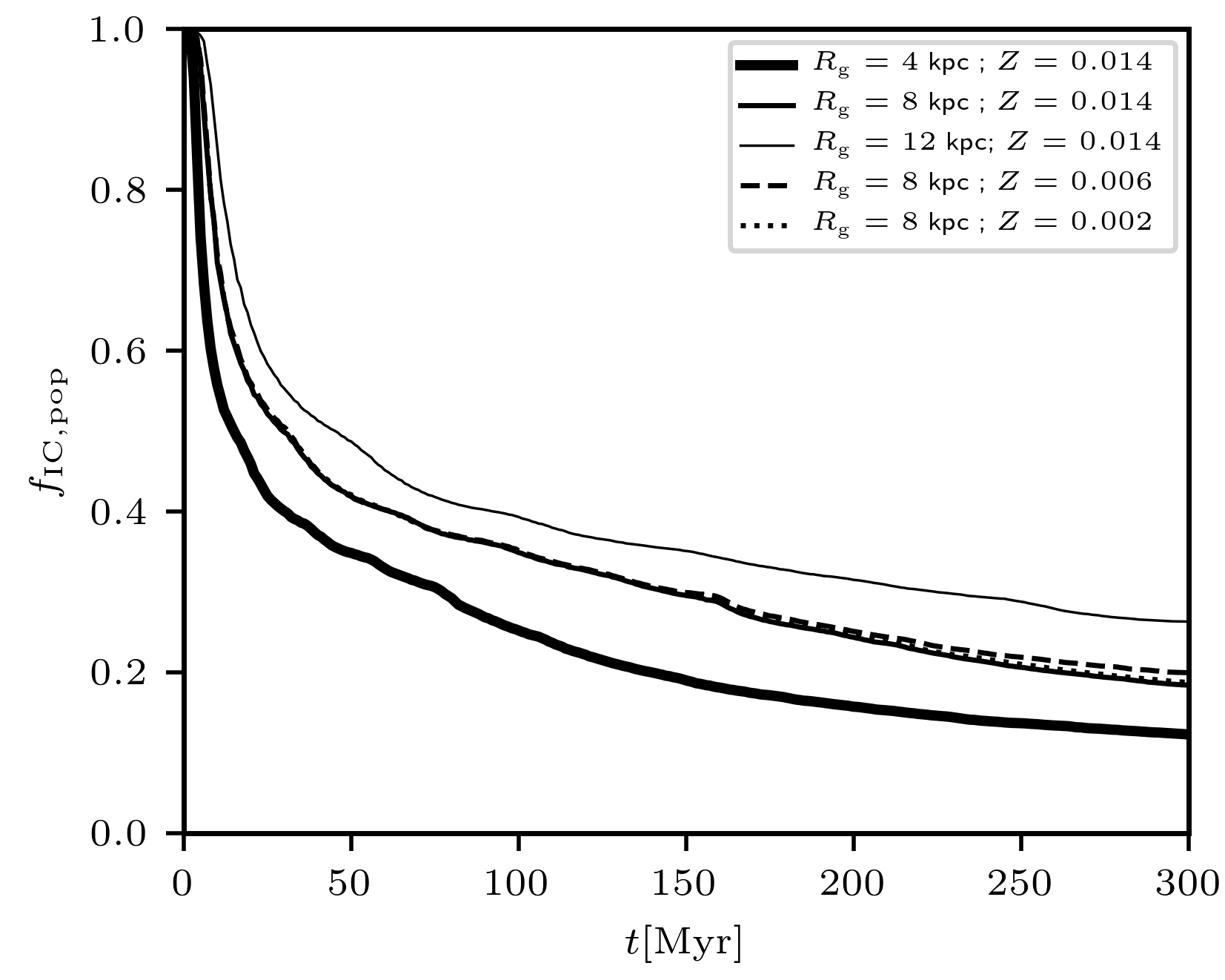}
\caption{The fraction of all stars located in their parent cluster for the galaxy-wide ECMF of \eq{eicmf} and for a single star-burst. 
Clusters orbit the galaxy at
$R_{\rm g} = 4 \Kpc$, $R_{\rm g} = 8 \Kpc$, and $R_{\rm g} = 12 \Kpc$ (different thickness of solid lines).
These clusters have the Solar metallicity of $Z = 0.014$.
Models of $Z = 0.006$ and $Z = 0.002$ for $R_{\rm g} = 8 \Kpc$ are shown by the dashed and dotted lines.
}
\label{fpopulationIC}
\end{figure} \else \fi

We assume that embedded clusters are distributed according to the embedded cluster mass function (ECMF), 
\begin{equation}
\xi_{\rm ecl} (M_{\rm ecl}) \equiv \nder{N_{\rm ecl}}{M_{\rm ecl}} \propto M_{\rm ecl}^{-\beta}, 
\label{eicmf}
\end{equation}
where $\dd N_{\rm ecl}$ is the number of clusters in the mass bin $(M_{\rm ecl}, M_{\rm ecl} + \dd M_{\rm ecl})$, 
and the index $\beta$ is a parameter obtained from observations. 
Let a cluster of mass $M_{\rm ecl}$ have a property $q(M_{\rm ecl})$, which is normalised by the total number of stars 
($q$ can be for example $f_{\rm IC}$).
Then, for a coeval population of star clusters with mass range $(M_{\rm ecl, min}, M_{\rm ecl, max})$,
\begin{equation}
q_{\rm pop} = \frac{\int_{M_{\rm ecl, min}}^{M_{\rm ecl, max}} \xi_{\rm ecl} (M_{\rm ecl}) q(M_{\rm ecl}) M_{\rm ecl} \dd M_{\rm ecl}}
{\int_{M_{\rm ecl, min}}^{M_{\rm ecl, max}} \xi_{\rm ecl} (M_{\rm ecl}) M_{\rm ecl} \dd M_{\rm ecl}}. 
\label{qmean}
\end{equation}
We note that the integrals are mass weighted because the contribution of each mass bin $\dd M_{\rm ecl}$ is proportional to the total 
number of stars which formed within the mass bin, and thus in turn to the cluster mass $M_{\rm ecl}$.
In the following text, we refer to a quantity calculated by \eq{qmean} as galaxy-wide. 
In the Galaxy and nearby galaxies, $\beta \approx 2$ \citep{vdBergh1984,Whitmore1999,Larsen2002,Lada2003, Bik2003,FuenteMarcos2004,Gieles2006}, 
and the cluster mass range covers approximately an interval of $(30 \Msun, 10^4 \Msun)$ \citep{Johnson2017}. 

For an ECMF with these parameters, we obtain the time dependence of the galaxy wide fraction $f_{\rm IC, pop}$ of stars in clusters as 
shown in \reff{fpopulationIC}. 
We assume that all the clusters were formed in a single starburst, i.e. that they are coeval.
The plot is constructed for clusters orbiting at different galactocentric radii $R_{\rm g}$, and with different metallicities for $R_{\rm g} = 8 \Kpc$. 
For all models, $f_{\rm IC, pop}$ drops significantly during the first $50 \Myr$ of cluster evolution. 
This drop is dominated by the early gas expulsion, with the galactic tidal field being of secondary importance. 
At a given time, the value of $f_{\rm IC, pop}$ increases with increasing $R_{\rm g}$ as the tidal field of the galaxy becomes less important. 
In particular, at $200 \Myr$ and for $R_{\rm g} = 4 \Kpc$, $8 \Kpc$ and $12 \Kpc$,  $f_{\rm IC, pop} = 0.16$, $0.25$ and $0.32$, respectively. 

The models with subsolar metallicities of $Z = 0.002$ and $Z=0.006$ are shown by the dotted and dashed lines in \reff{fpopulationIC}, respectively. 
They indicate that $f_{\rm IC, pop}$ is practically independent of the metallicity of the clusters. 

\reff{fIndividualStStars} shows the time dependence of the galaxy-wide fraction $f_{\rm IC, pop}(m)$ of stars in clusters for stars 
of different masses $m$ (or equivalently different spectral types). 
The figure demonstrates that the clusters mass segregate because the probability to find more massive stars (with the exception of O stars) 
in clusters at a given time is larger than to find there lower mass stars. 
The mass segregation is dynamical because these models are free of primordial mass segregation. 
At a given time, the fraction $f_{\rm IC, pop}(m)$ monotonically increases with stellar mass apart from the O stars, because many O stars are ejected in 
close three- and many- body interactions producing runaway stars (e.g. \citealt[][]{Fujii2011,Tanikawa2012,Perets2012,Oh2016}). 
At the age of $200 \Myr$, only $19$\% of M stars are located in clusters, while $32$\% of late B stars are located in clusters. 

\iffigscl
\begin{figure}
\includegraphics[width=\columnwidth]{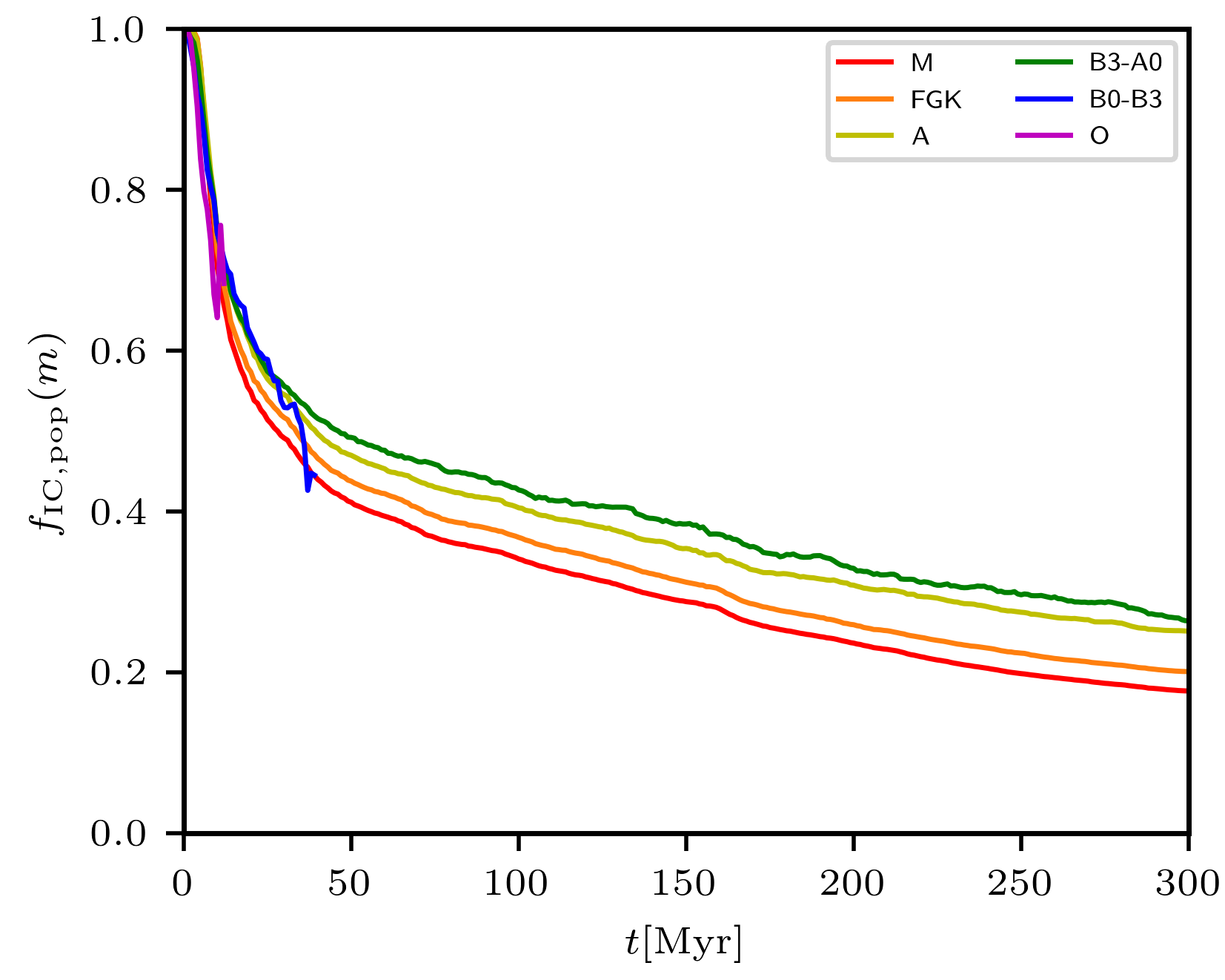}
\caption{The time dependence of the fraction of M (red line), F-K (orange), A (yellow), late B (green), early B (blue) and O (magenta) stars 
located in clusters for clusters with $R_{\rm g} = 8 \Kpc$ and $Z = 0.014$. 
The clusters are assumed to form with the ECMF of \eq{eicmf} with $\beta = 2$, $M_{\rm ecl, min} = 30 \Msun$, and $M_{\rm ecl, max} = 10^4 \Msun$. 
At a given time, $f_{\rm IC, pop}(m)$ increases with stellar mass (apart from O stars), which is a result of dynamically induced mass segregation.
O stars are slightly depleted in clusters because of their dynamical ejections.
}
\label{fIndividualStStars}
\end{figure} \else \fi

\subsection{Fraction of stars in clusters in the context of the IGIMF in a single starburst}

\label{ssIGIMF}

We explore the behaviour of $f_{\rm IC, pop}$ in the 
IGIMF framework, 
which was proposed by \citet{Kroupa2003b} and further developed and refined by \citet{Weidner2013,Yan2017} and \citet{Jerabkova2018}. 
According to the IGIMF theory, the upper mass limit for star clusters $M_{\rm ecl, max}$ increases, and $\beta$ decreases with increasing the SFR, so  
environments with larger SFR tend to form more massive clusters, and these clusters are more abundant relative to the lower mass clusters. 
The SFR is taken from the whole galaxy.
Thus, the IGIMF theory allows the calculation of a galaxy-wide stellar population
assuming all stars form in embedded star clusters. 
It successfully couples the molecular 
cloud-core scale to galaxy-wide properties such as the star formation rate.

We study galaxies with SFRs of $0.01 \Mspy$, $0.1 \Mspy$, $1 \Mspy$, $10 \Mspy$ and $100 \Mspy$. 
For each of these galaxies, we determine the mass of the most massive cluster according 
to eq. 6 of \citet{Weidner2004} (see also \citealt{Randriamanakoto2013}), 
which reads
\begin{equation}
M_{\rm ecl, max} = 8.5 \times 10^4 \Msun (\sfr/(\Mspy))^{0.75}.
\label{eClMax}
\end{equation}
Since galaxies with $\sfr \gtrsim 0.1 \Mspy$ form clusters more massive than we can simulate easily (a $100 \Mspy$ galaxy has $M_{\rm ecl,max} = 2.7 \times 10^6 \Msun$
 according to \eq{eClMax}, which is beyond the capability of even the most dedicated current software; \citealt{Wang2015,Wang2020}), 
we extrapolate 
\footnote{From the requirement that the extrapolation function tends to $1$ for $M_{\rm ecl} \to \infty$ and to $0$ for $M_{\rm ecl} \to 0$, 
we adopt the extrapolation function in the form of 
\begin{equation}
y(\log_{10}(M_{\rm ecl})) = \frac{1}{\pi} \left\{ \frac{\pi}{2} + \arctan(a(\rm{log}_{10}(M_{\rm ecl})) - b) \right\}, 
\label{eExtrapol}
\end{equation}
where $a$ and $b$ are the unknowns.
}
$f_{\rm IC} (M_{\rm ecl})$ from the five most massive clusters in our simulations towards higher mass at any time $t$. 
The least massive clusters dissolve early, so we did not calculate extra models for $M_{\rm ecl} < 50 \Msun$, but extrapolate the values of 
$f_{\rm IC} (M_{\rm ecl})$ from the five least massive clusters in our simulations towards lower cluster mass. 

\iffigscl
\begin{figure}
\includegraphics[width=\columnwidth]{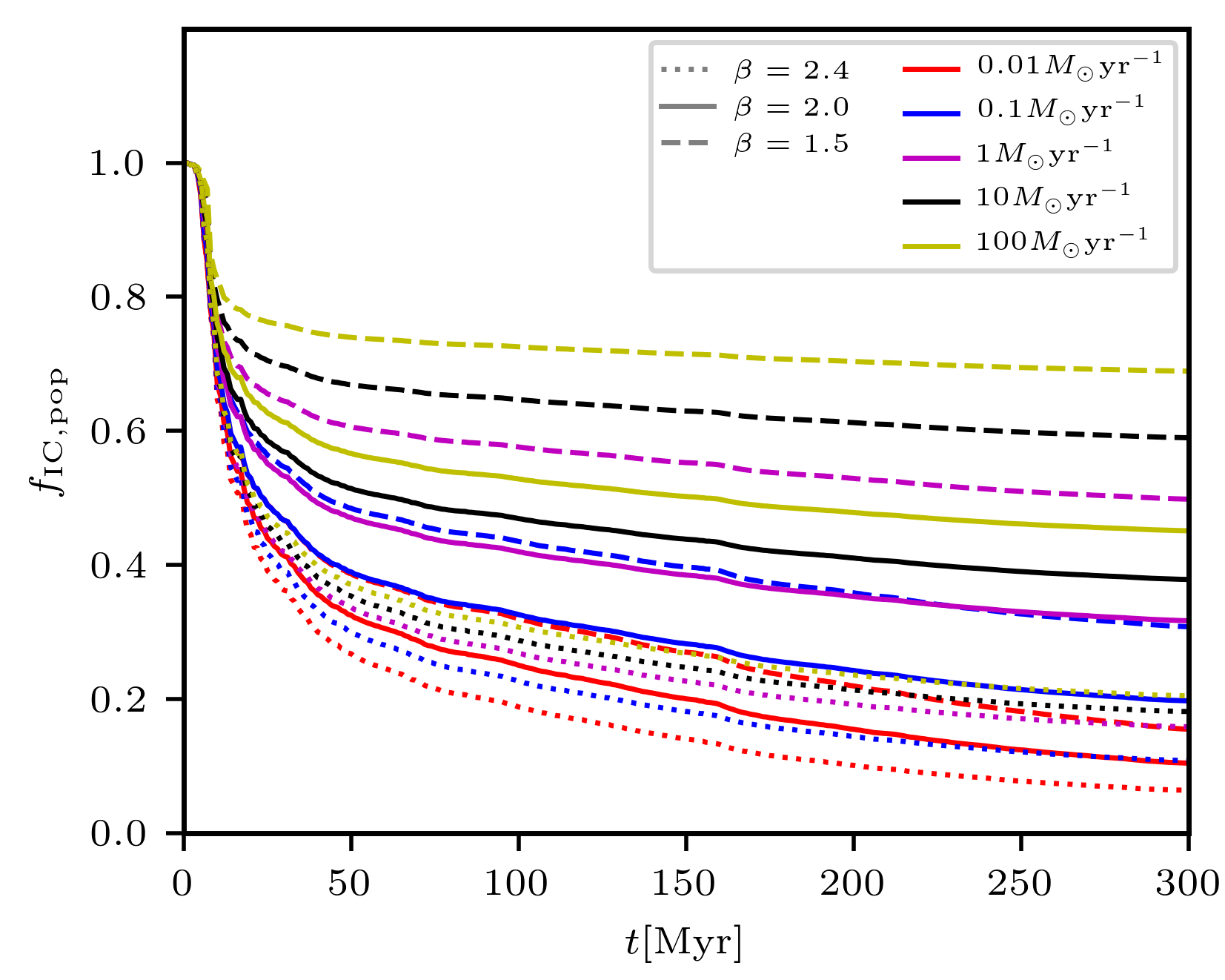}
\caption{The time dependence of the galaxy-wide $f_{\rm IC,pop}$ for galaxies with SFRs of $0.01 \Mspy$, $0.1 \Mspy$, $1 \Mspy$, $10 \Mspy$ and $100 \Mspy$ 
(indicated by line colour) according to the IGIMF theory. 
The clusters have a metallicity of $Z = 0.014$ and orbit the galaxy at $R_{\rm g} = 8 \Kpc$.
The cluster mass range spans from $5 \Msun$ to $M_{\rm ecl,max} (\sfr)$, which is given by \eq{eClMax}.
For each value of the SFR, we consider $\beta = 1.5$ (dashed lines), $\beta = 2$ (solid lines), and $\beta = 2.4$ (dotted lines).
}
\label{fpopulationIC_betaSFR}
\end{figure} \else \fi

The galaxy-wide fraction of stars in clusters as calculated by \eq{qmean} is shown in \reff{fpopulationIC_betaSFR}. 
We assume that all the clusters were formed at the same time.
In order to be consistent with the IGIMF framework, we set the ranges of the ECMF to be $M_{\rm ecl,min} = 5 \Msun$ and $M_{\rm ecl,max}$ according to \eq{eClMax}, 
which is different from the cluster population studied in \refs{ssGalaxyWide}. 
Here, we take $\beta$ and the SFR as independent variables to explore the admitted uncertainty in $f_{\rm IC,pop}$; 
although in the strict sense of the IGIMF theory, $\beta$ is a function of the SFR. 
According to the IGIMF theory, $\beta$ ranges from $2.2$ to $1.8$ \citep{Weidner2004,Jerabkova2018} in the considered range 
of the SFR of $\sfr = 0.01 \Mspy$ to $\sfr = 100 \Mspy$. 
This is well within the extreme values of $\beta$ of $1.5$ and $2.4$ considered in this work.

For any $\beta$, at a given time, $f_{\rm IC,pop}$ increases with increasing $\sfr$ because of the increase of $M_{\rm ecl,max}$. 
Particularly, for $\beta = 2$ at $t = 200 \Myr$, $f_{\rm IC,pop} = 0.15$, $0.35$ and $0.48$ for SFR of $0.01 \Mspy$, $1 \Mspy$ and $100 \Mspy$, respectively. 
The ECMF of a shallower slope of $\beta = 1.5$ (dashed lines in \reff{fpopulationIC_betaSFR}) produces more massive clusters relative to the lower mass ones, 
these clusters dissolve at a slower rate, which results in an increase of $f_{\rm IC,pop}$ at a given time. 
In contrast, ECMFs of a steeper slope ($\beta = 2.4$; dotted lines) form more low mass clusters, which are easily disrupted, 
which results in a lower value of $f_{\rm IC,pop}$. 
The likely observed fraction of stars in clusters within a galaxy of $\rm{SFR} = 1 \Msun \rm{yr}^{-1}$ (Milky Way-like SFR)
in the IGIMF theory (but with $\beta$ taken as a free parameter) is shown in the right panel of \reff{fobserved}.

\subsection{Fraction of stars in clusters for a galaxy with a constant rate of star formation}

\label{ssContSFR}

\begin{table*}
\begin{tabular}{ccccccc}
$\beta$ & $\sfr$ & $M_{\rm ecl,min}$ & $M_{\rm ecl,max}$ & Spectral & $R_{\rm g}$ & $\overline{f}_{\rm IC,pop} (t < 300 \Myr)$ \\
 & [$\Mspy$]  & [$\Msun$] & [$\Msun$] & type & [$\Kpc$] & \\
\hline
  2.0 & - & 30 & 1.0$\times$10$^{4}$ & all & 4 & 0.26 \\
  2.0 & - & 30 & 1.0$\times$10$^{4}$ & all & 8 & 0.33 \\
  2.0 & - & 30 & 1.0$\times$10$^{4}$ & all & 12 & 0.41 \\
\hline
  2.0 & - & 30 & 2.0$\times$10$^{4}$ & all & 4 & 0.36 \\
  2.0 & - & 30 & 4.0$\times$10$^{3}$ & all & 8 & 0.30 \\
  2.0 & - & 30 & 1.0$\times$10$^{3}$ & all & 12 & 0.29 \\
\hline
  2.0 & - & 30 & 1.0$\times$10$^{4}$ & O-B3 & 8 & 0.83 \\
  2.0 & - & 30 & 1.0$\times$10$^{4}$ & B0-B3 & 8 & 0.67 \\
  2.0 & - & 30 & 1.0$\times$10$^{4}$ & B3-A0 & 8 & 0.41 \\
  2.0 & - & 30 & 1.0$\times$10$^{4}$ & A & 8 & 0.39 \\
  2.0 & - & 30 & 1.0$\times$10$^{4}$ & F-K & 8 & 0.35 \\
  2.0 & - & 30 & 1.0$\times$10$^{4}$ & M & 8 & 0.32 \\
\hline
   1.5 &   0.01 & 5 & 2.7$\times$10$^{3}$ & all & 8 &   0.39  \\
   1.5 &   0.10 & 5 & 1.5$\times$10$^{4}$ & all & 8 &   0.55  \\
   1.5 &   1.00 & 5 & 8.5$\times$10$^{4}$ & all & 8 &   0.72  \\
   1.5 &  10.00 & 5 & 4.8$\times$10$^{5}$ & all & 8 &   0.78  \\
   1.5 & 100.00 & 5 & 2.7$\times$10$^{6}$ & all & 8 &   0.84  \\
   2.0 &   0.01 & 5 & 2.7$\times$10$^{3}$ & all & 8 &   0.24  \\
   2.0 &   0.10 & 5 & 1.5$\times$10$^{4}$ & all & 8 &   0.32  \\
   2.0 &   1.00 & 5 & 8.5$\times$10$^{4}$ & all & 8 &   0.42  \\
   2.0 &  10.00 & 5 & 4.8$\times$10$^{5}$ & all & 8 &   0.47  \\
   2.0 & 100.00 & 5 & 2.7$\times$10$^{6}$ & all & 8 &   0.53  \\
   2.4 &   0.01 & 5 & 2.7$\times$10$^{3}$ & all & 8 &   0.16  \\
   2.4 &   0.10 & 5 & 1.5$\times$10$^{4}$ & all & 8 &   0.17  \\
   2.4 &   1.00 & 5 & 8.5$\times$10$^{4}$ & all & 8 &   0.18  \\
   2.4 &  10.00 & 5 & 4.8$\times$10$^{5}$ & all & 8 &   0.19  \\
   2.4 & 100.00 & 5 & 2.7$\times$10$^{6}$ & all & 8 &   0.19
\end{tabular}
\caption{
The fraction of stars in clusters $\overline{f}_{\rm IC,pop} (t < 300 \Myr)$ for a time independent SFR for stars younger than $300 \Myr$. 
The data are divided into four groups as indicated by the horizontal lines. 
The first group of three rows corresponds to the models of \refs{ssGalaxyWide} with fixed $M_{\rm ecl,max}$, 
which are calculated at three different $R_{\rm g}$.
The second group is for the same models but assuming $M_{\rm ecl,max}$ dependent on the galactocentric radius according to \citet{PflammAltenburg2008}.
The third group of six rows lists the fraction of stars in clusters according to their spectral type for the models with $R_{\rm g} = 8 \Kpc$. 
The last group represents the IGIMF models with different ECMF slopes $\beta$ and SFRs (models of \refs{ssIGIMF}). 
The upper cluster mass limit $M_{\rm ecl,max}$ is a function of the assumed SFR \eqp{eClMax}. 
}
\label{tbetaSFR}
\end{table*}

In the previous two sections, Sect. \ref{ssGalaxyWide} and \ref{ssIGIMF}, we studied $f_{\rm IC,pop}$ 
for star clusters which all formed in a single starburst. 
Here, we consider the opposite scenario where star formation proceeds continuously with the SFR constant in time. 
The particular value of the SFR is not important as long as it allows us to fully sample the upper limit of the ECMF, 
because of the normalisation to the total number of stars formed.
The results are relevant to stars younger than $300 \Myr$ only because of the duration of our simulations. 
Such a star formation history produces a time-independent fraction $\overline{f}_{\rm IC,pop} (t < 300 \Myr)$ of stars in clusters, which is listed in \reft{tbetaSFR}.

The first three lines of the table feature star clusters with $(M_{\rm ecl,min}, M_{\rm ecl,max}) = (30 \Msun, 10^4 \Msun)$ and $\beta = 2$. 
In these models, we use the same cluster mass range at all the galactocentric radii $R_{\rm g}$ to study the influence of the tidal field strength. 
For these clusters, $\overline{f}_{\rm IC,pop} (t < 300 \Myr)$ increases with increasing $R_{\rm g}$ from $0.26$ for $R_{\rm g} = 4\Kpc$ to $0.41$ for $R_{\rm g} = 12\Kpc$.

There is evidence that the mass of the 
most massive cluster decreases with the local gas column density \citep{PflammAltenburg2008}. 
In disc galaxies, the dependence can be approximated as $\propto \exp(-3R_{\rm g}/2R_{\rm d})$, where $R_{\rm d}$ is the disc scale-length. 
Taking this dependence into account (with $R_{\rm d} = 4 \Kpc$), we obtain $\overline{f}_{\rm IC,pop} (t < 300 \Myr)$ decreasing with 
$R_{\rm g}$ from $0.36$ for $R_{\rm g} = 4\Kpc$ to $0.29$ for $R_{\rm g} = 12\Kpc$ (second group of models in \reft{tbetaSFR}). 
Thus, the decreasing $M_{\rm ecl,max}$ with $R_{\rm g}$ acts in the opposite way than the decreasing tidal field strength (as discussed in the previous paragraph), 
and it has more impact on $\overline{f}_{\rm IC,pop}$ than the tidal field.

The third group of records in \reft{tbetaSFR} lists the time-averaged fractions 
of stars in clusters for stars in six different mass bins for clusters with $R_{\rm g} = 8\Kpc$ and $M_{\rm ecl,max} = 10^4 \Msun$. 
The value of $\overline{f}_{\rm IC,pop}(m, t < 300 \Myr)$ increases monotonically with stellar mass from $0.32$ for M stars to $0.41$ for late B (B3-A0) stars. 
Stars more massive than that have substantially higher $\overline{f}_{\rm IC,pop}(m, t < 300 \Myr)$ (reaching $0.83$ for O and early B stars) 
not only as the result of dynamical mass segregation, but also 
because they are present only in the youngest clusters, which have had not enough time to dissolve.

The last 15 lines of \reft{tbetaSFR} represent the results with the cluster mass limits $(M_{\rm ecl,min}, M_{\rm ecl,max})$ set according 
to the IGIMF theory (\refs{ssIGIMF}). 
The Table shows the same trend as we encountered in \refs{ssIGIMF}:  $\overline{f}_{\rm IC,pop} (t < 300 \Myr)$ increases with increasing $\sfr$ and decreasing $\beta$. 
For models with a top-heavy ECMF ($\beta = 1.5$), $\overline{f}_{\rm IC,pop} (t < 300 \Myr)$ strongly depends on the $\sfr$: 
$\overline{f}_{\rm IC,pop} (t < 300 \Myr)$ increases by more than a factor of $2$ from $\overline{f}_{\rm IC,pop} (t < 300 \Myr) = 0.39$ for $\sfr = 0.01 \Mspy$ to 
$\overline{f}_{\rm IC,pop} (t < 300 \Myr) = 0.84$ for $\sfr = 100 \Mspy$.
On the other hand, models with a top-light ECMF ($\beta = 2.4$) have $\overline{f}_{\rm IC,pop} (t < 300 \Myr)$ almost independent of $\sfr$ because such a steep 
$\beta$ ensures that there are very few massive star clusters, which would contribute to a higher $\overline{f}_{\rm IC,pop} (t < 300 \Myr)$. 

The model of $\beta = 2$ and $\sfr = 1 \Mspy$ is close to the Galaxy in the IGIMF framework, which includes the cluster population of the whole Galaxy. 
This model provides $\overline{f}_{\rm IC,pop} (t < 300 \Myr) = 0.42$. 
Since the mass of the most massive cluster decreases with the local gas column density \citep{PflammAltenburg2008}, 
the majority of the most massive star clusters in the Galaxy are located near the Galactic centre or tips of the bar \citep{Zwart2010}, 
and the environment in the vicinity of the Sun hosts clusters with masses only up to $M_{\rm ecl,max} \approx 4000 \Msun$. 
This cluster mass limit provides $\overline{f}_{\rm IC,pop} (t < 300 \Myr) = 0.26$, which is by $\approx 40$\% lower than 
what we obtain for the Galaxy as a whole.

\subsection{Comparison to observations and to the current theoretical model}

\label{ssCombObs}

\iffigscl
\begin{figure}
\includegraphics[width=\columnwidth]{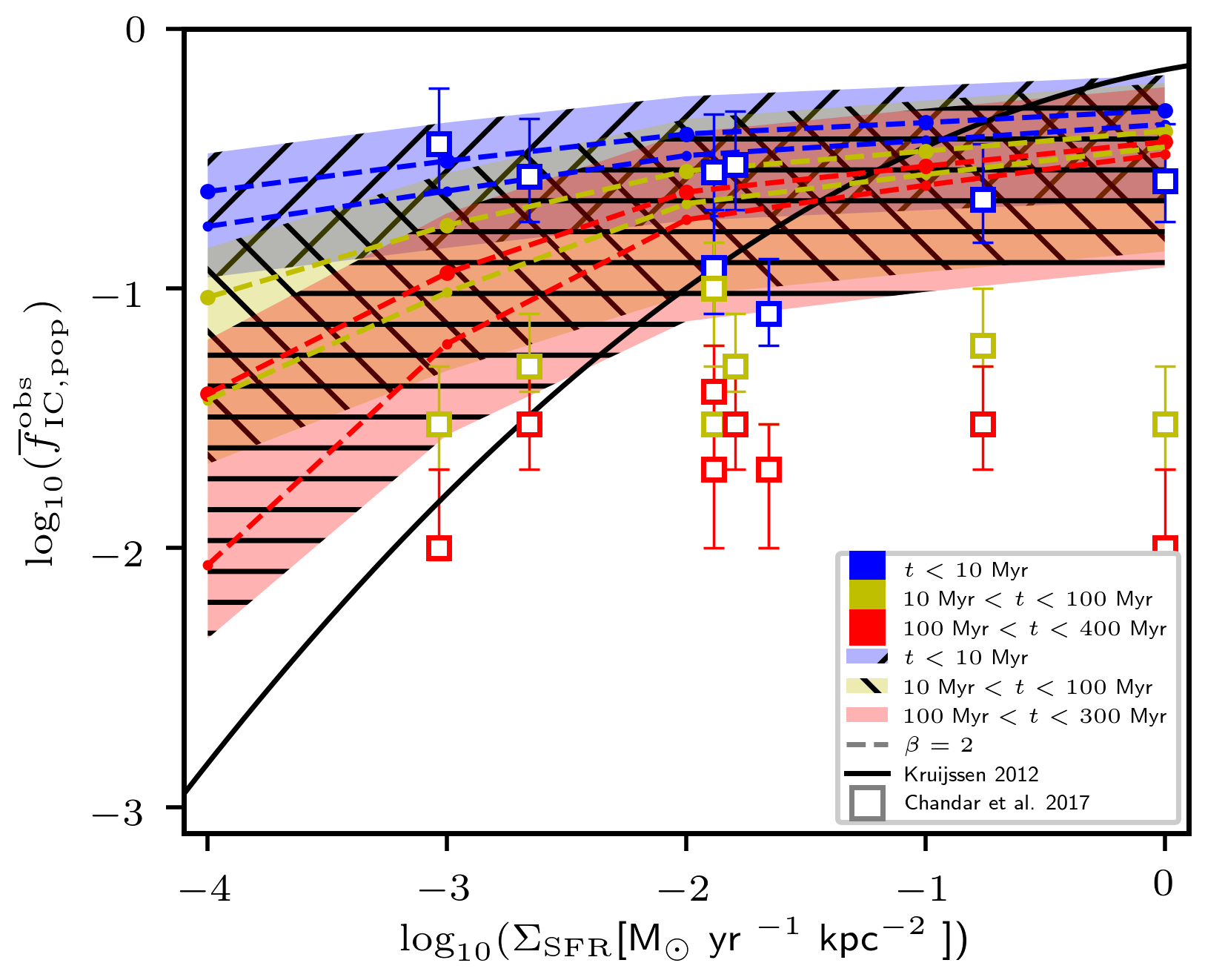}
\caption{
The likely detected fraction of stars in clusters of age $t < 10 \Myr$ (blue area and blue lines), $10 \Myr < t < 100 \Myr$ (yellow area and yellow lines),
and $100 \Myr < t < 300 \Myr$ (red area and red lines) as a function of $\Sigma_{\rm SFR}$. 
Apart from indicating age by a colour, age is also shown by a hatch style. 
The coloured area is bordered by two extreme models: the lower one has $\beta = 2.2$ and $r_{\rm search} = 2 \Pc$, the upper one has 
$\beta = 1.8$ and $r_{\rm search} = 5 \Pc$. 
The models for $\beta = 2.0$ and $r_{\rm search} = 2 \Pc$ and $5 \Pc$ are shown by the dashed lines. 
The plot shows how $\overline{f}^{\rm obs}_{\rm IC,pop}$ increases with the $\Sigma_{\rm SFR}$, and how it decreases with the cluster age.
The observational data due to \citet{Chandar2017}, which aim consistently at galaxies both with low SFR and high SFR, are 
shown by empty squares with their errorbars. 
Our results are obtained at almost the same time intervals as the observational data (indicated by the colours). 
The semi-analytic estimate provided by the \GamSig model is shown by the black solid line. 
Note its disagreement with the observational data for the clusters of any age interval.
}
\label{fsigmaSFR}
\end{figure} \else \fi

The observations of \citet{Chandar2017} present results for eight galaxies. 
We focus particularly on Small and Large Magellanic Clouds, because they differ most from the \GamSig model (the two data 
points with the lowest $\Sigma_{\rm SFR}$ in \reff{fsigmaSFR}) at their youngest age. 
These data points utilise the cluster catalogue of \citet{Hunter2003}, which is based on visual inspection of cluster candidates from previous 
catalogues. 
\citet{Hunter2003} take a candidate object as a cluster if it is distinguishable from the background galaxy within a 
circle of radius 23" ($5.5 \Pc$ at the distance of $50 \Kpc$), and resolved at least to several stars. 
This makes our simulations difficult to be directly compared with observations because we cannot automatically identify star clusters in our simulations 
based on exactly defined criteria. 

To provide an order of magnitude estimate, we place a circle of radius $r_{\rm search}$ at the cluster density centre, and 
count the number of stars located within the projected circle. 
If at least 10 stars of spectral type A and earlier are found, we consider the cluster to be detected, and we count the stars 
within radius $r_{\rm search}$ as the cluster population, while the rest of the stars count to the field population. 
Otherwise, we do not consider the cluster to be detected and count all stars in it as the field population. 
To estimate the sensitivity of this method, we use two radii: $2 \Pc$ and $5 \Pc$, the latter one is close to the radius $\approx 5.5 \Pc$ adopted by \citet{Hunter2003}.

\reff{fsigmaSFR} compares the fraction of stars detected by the method described above in our simulations (hatched areas and dashed lines) 
with the observational data due to \citet{Chandar2017} (empty squares with error bars). 
The SFR is related to the $\Sigma_{\rm SFR}$ on the assumption that each galaxy has a star formation area of $100 \Kpc^2$; this order of magnitude estimate is 
compatible with the galaxies studied by \citet{Chandar2017} (cf. their table 1).
We evaluate $\overline{f}^{\rm obs}_{\rm IC,pop}$ at the same time intervals as the observational data except the oldest age interval, 
which we take from 100 Myr to 300 Myr, while the observations take it from 100 Myr to 400 Myr. 
The hatched areas cover the range of $\beta \in (1.8, 2.2)$ and $r_{\rm search}$ between $2$ and $5 \Pc$, while 
the dashed lines show the value of $\overline{f}^{\rm obs}_{\rm IC,pop}$ for $\beta = 2$ and $r_{\rm search} = 2 \Pc$ and $5 \Pc$. 
The youngest clusters (blue symbols and lines; age $< 10 \Myr$) have $\overline{f}^{\rm obs}_{\rm IC,pop}$ comparable to observations, while older clusters 
have $\overline{f}^{\rm obs}_{\rm IC,pop}$ higher than in observations. 
Thus, the rough approximation to 'synthetic observations' is in agreement with the original hypothesis that all stars originate in gravitationally bound embedded clusters. 
Gas expulsion not only unbinds many stars from the clusters, but also decreases 
their density so that many of the youngest clusters cannot be detected, an idea originally proposed by \citet{Lueghausen2012}. 
The overestimate of $\overline{f}_{\rm IC,pop}$ in older clusters can be caused by neglecting the interaction of clusters with molecular clouds 
in our models, which probably plays an important role in cluster dissolution \citep{Terlevich1987,Lamers2006,Chandar2010,Jerabkova2021}.


Only the youngest objects can be compared with the \GamSig theory because the theory does not consider 
any process which releases stars after the clusters emerge from their natal clouds, which occurs at the age of several Myr. 
This constraints the comparison to the youngest objects ($10 \Myr$; blue lines and squares in \reff{fsigmaSFR}), 
which are the least impacted by dynamical evolution. 
Thus, $\Gamma \approx \overline{f}_{\rm IC,pop} (t < 10 \Myr)$.
For low SFRs, \GamI models predict substantially larger values of $\overline{f}^{\rm obs}_{\rm IC,pop}(t < 10 \Myr)$ in contrast to the \GamSig model:
for $\Sigma_{\rm SFR} = 10^{-4} \Msun \rm{yr}^{-1} \Kpc^{-2}$, the admitted values of $\overline{f}^{\rm obs}_{\rm IC,pop}(t < 10 \Myr)$ 
in \GamI models range from $0.07$ to $0.45$, 
while the \GamSig model predicts $\Gamma = 0.0015$. 
\footnote{We note that we compare the 'observed' fraction of stars in clusters in the \GamI model (i.e. $\overline{f}^{\rm obs}_{\rm IC,pop}(t < 10 \Myr)$) 
with the physical fraction (i.e. $\Gamma$) of the \GamSig model. 
However, the observed fraction of stars in clusters is always lower than the physical fraction 
(i.e., $\overline{f}^{\rm obs}_{\rm IC,pop}(t < 10 \Myr) < \overline{f}_{\rm IC,pop}(t < 10 \Myr)$) 
because of incompleteness, which makes the difference with the \GamSig model even more pronounced.
}
For high SFRs, the \GamI model predicts lower $\overline{f}^{\rm obs}_{\rm IC,pop}(t < 10 \Myr)$ in contrast to the \GamSig model. 

In the \GamSig model as well as in the \GamI model, $\Gamma$ increases with $\Sigma_{\rm SFR}$. 
However, the reason for this behaviour is very different. 
In the \GamSig model, it is the combined effect of the increasing $\sfe(x)$ and $\gamma(x)$ with $x$, where high values of $x$ are reached 
as the ISM probability distribution function widens with increasing $\rho_{\rm ISM}$ (via the ISM Mach number). 
In contrast, in the \GamI model, the increase is due to the expansion and partial dissolution of clusters as the result of gas expulsion, which 
has a lesser impact on more massive clusters, which according to the IGIMF theory form only in galaxies with a sufficiently high $\Sigma_{\rm SFR}$.
Another difference is the much weaker dependence of $\Gamma$ on $\Sigma_{\rm SFR}$, which 
was noted by \citet{Chandar2017}, and which is reproduced in the \GamI model.

\section{Discussion}

\label{sDiscuss}

\subsection{The influence of the initial conditions of star clusters}

\label{ssInfluenceICs}

To calculate the present models, we adopt particular initial conditions, which are described in \refs{sssStandardModels}. 
However, both observational and theoretical evidence indicate that the initial conditions for star cluster formation are not fully understood. 
For example, the value of the SFE, initial cluster radii, the time-scale of gas expulsion and the degree of primordial mass segregation are subject to ongoing research 
\citep[e.g.][]{Er2013,Kuhn2014,Banerjee2015,Megeath2016,Spera2017,Dominguez2017,Pfalzner2019,Kuhn2019,Dib2019,Pang2020}. 
Some of the cluster initial conditions are interrelated: for example, embedded star clusters cannot form with sub-parsec scales and SFE close to $100$\% 
because they would not be able to expand to their observed sizes \citep{Banerjee2017}.

\subsubsection{Initial cluster radius and the SFE}

\label{sssInfluenceRadSFE}

To study the possible influence of the initial cluster radius and the SFE, we perform additional models 
with clusters of slightly larger radii (by at most a factor of $3$; "A" models) and clusters of a slightly lower SFE of $0.25$ ("F" models) 
(see \refs{sssAdditionalModels} for the model descriptions). 
The time dependence of $f_{\rm IC} (M_{\rm ecl})$ for clusters of four different masses is shown in \reff{fInfluenceRadSFE}.
In comparison to the standard models ("S" models), both "A" and "F" models release stars to the field earlier, and have a faster cluster dissolution. 
Also, the most massive clusters in our sample ($6400 \Msun$) have lost most of their stars already by $30 \Myr$ in "A" and "F" models, 
while the clusters are relatively unaffected in the standard models even at $300 \Myr$. 

For a Milky Way-like galaxy with constant star formation rate and star formation parameters $\beta = 2$ and $\sfr = 1 \Mspy$, "A" and "F" models 
have only $\overline{f}_{\rm IC,pop} (t < 300 \Myr) = 0.11$ and $0.06$, respectively, while it is $0.42$ in standard models. 
The difference of $f_{\rm IC,pop}$ between the models is small in the youngest clusters, but it gets pronounced with time. 
Model "A" has the interesting property of $f_{\rm IC,pop}(M_{\rm ecl})$ being almost independent of the cluster mass by $\approx 150 \Myr$ 
for the cluster mass range spanning more than one order of magnitude from $\approx 400 \Msun$ to $\approx 6400 \Msun$. 
This property is not seen in models "F" and "S", which have a smaller cluster radii. 
This behaviour is in line with the observations indicating that the dissolution rate of star clusters is independent of their initial mass 
\citep[e.g.][]{Lada2003,Fall2005,Whitmore2007}, 
but such an explanation is only hypothetical for now given the model uncertainties and the uncertain role of molecular clouds in cluster dissolution.

The physical reason for the faster release of stars from clusters in the "A" models is due to gas expulsion occurring more impulsively, i.e. the ratio 
between the gas expulsion time-scale and the cluster crossing time, $\tau_{\rm M}/t_{\rm cross}$, being smaller. 
Since $\tau_{\rm M}/t_{\rm cross} \propto \sqrt{M_{\rm ecl}/a_{\rm ecl}}$ and $a_{\rm ecl} \propto M_{\rm ecl}^{1/3}$, the difference between 
the "A" and "S" models increases with increasing cluster mass. 
The impact of gas expulsion increases strongly when $\tau_{\rm M}/t_{\rm cross}$ becomes smaller than unity \citep{Baumgardt2007}, 
which is the case of all clusters in the "A" models. 
The most massive clusters ($M_{\rm ecl} = 6400 \Msun$) have $\tau_{\rm M}/t_{\rm cross} = 1$ for "S" models, while it is $0.63$ for "A" models. 
Because of the monotonic increase of the ratio $\tau_{\rm M}/t_{\rm cross}$ with $M_{\rm ecl}$ (as $M_{\rm ecl}^{1/3}$ for "A" models), 
clusters with $M_{\rm ecl} < 6400 \Msun$ have $\tau_{\rm M}/t_{\rm cross} < 0.63$, resulting in their rapid dissolution. 
The reason for the faster release of stars from clusters in "F" models is due to their lower SFE \citep{Baumgardt2007}. 

\iffigscl
\begin{figure}
\includegraphics[width=\columnwidth]{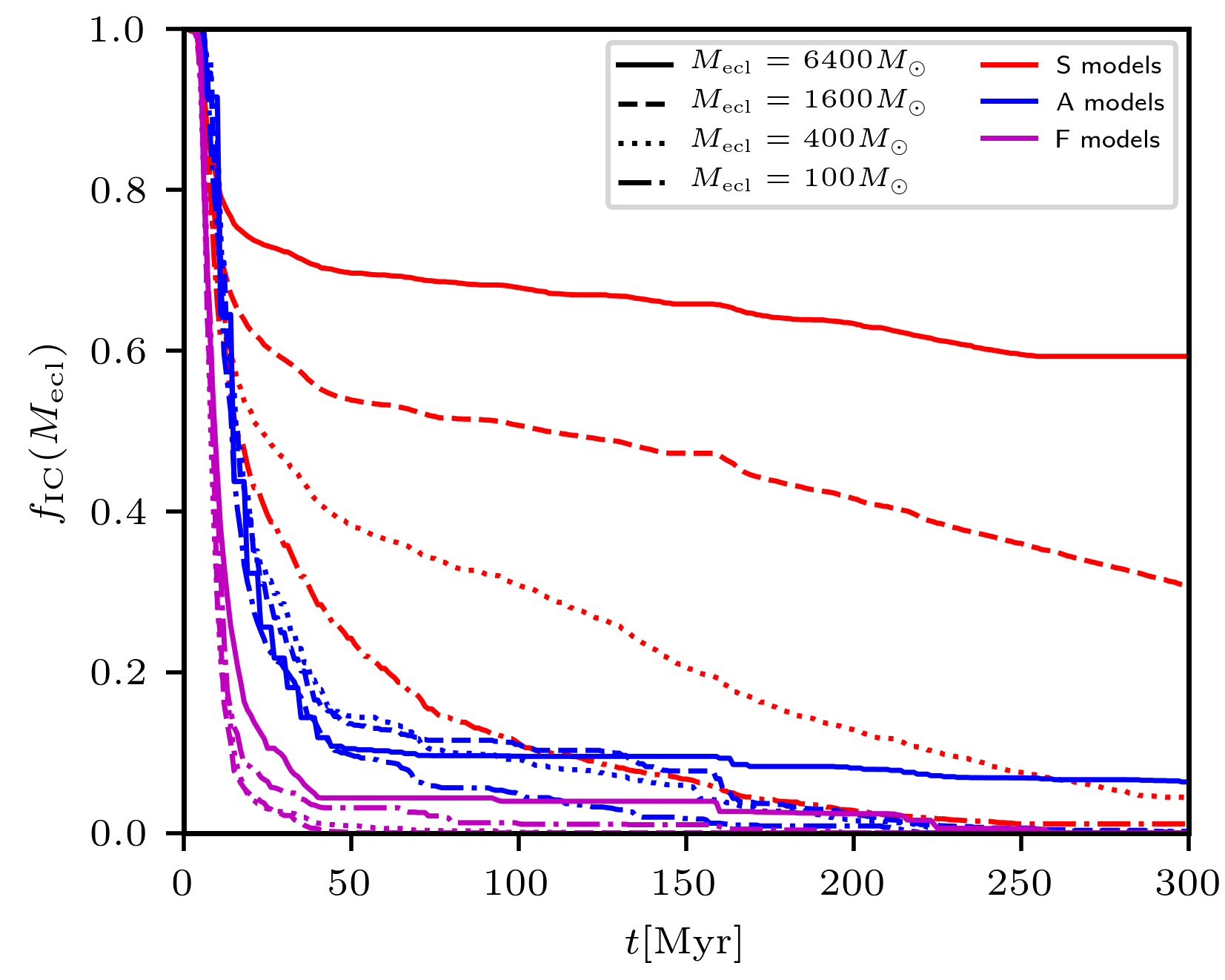}
\caption{
The fraction of stars located in clusters of an initial mass $M_{\rm ecl}$ for "A" models (blue lines) and "F" models (magenta lines). 
For comparison, standard models are shown by red lines. 
"A" and "F" models result in substantially lower fraction of stars in clusters of all considered masses at $t \gtrsim 20 \Myr$. 
}
\label{fInfluenceRadSFE}
\end{figure} \else \fi

\iffigscl
\begin{figure}
\includegraphics[width=\columnwidth]{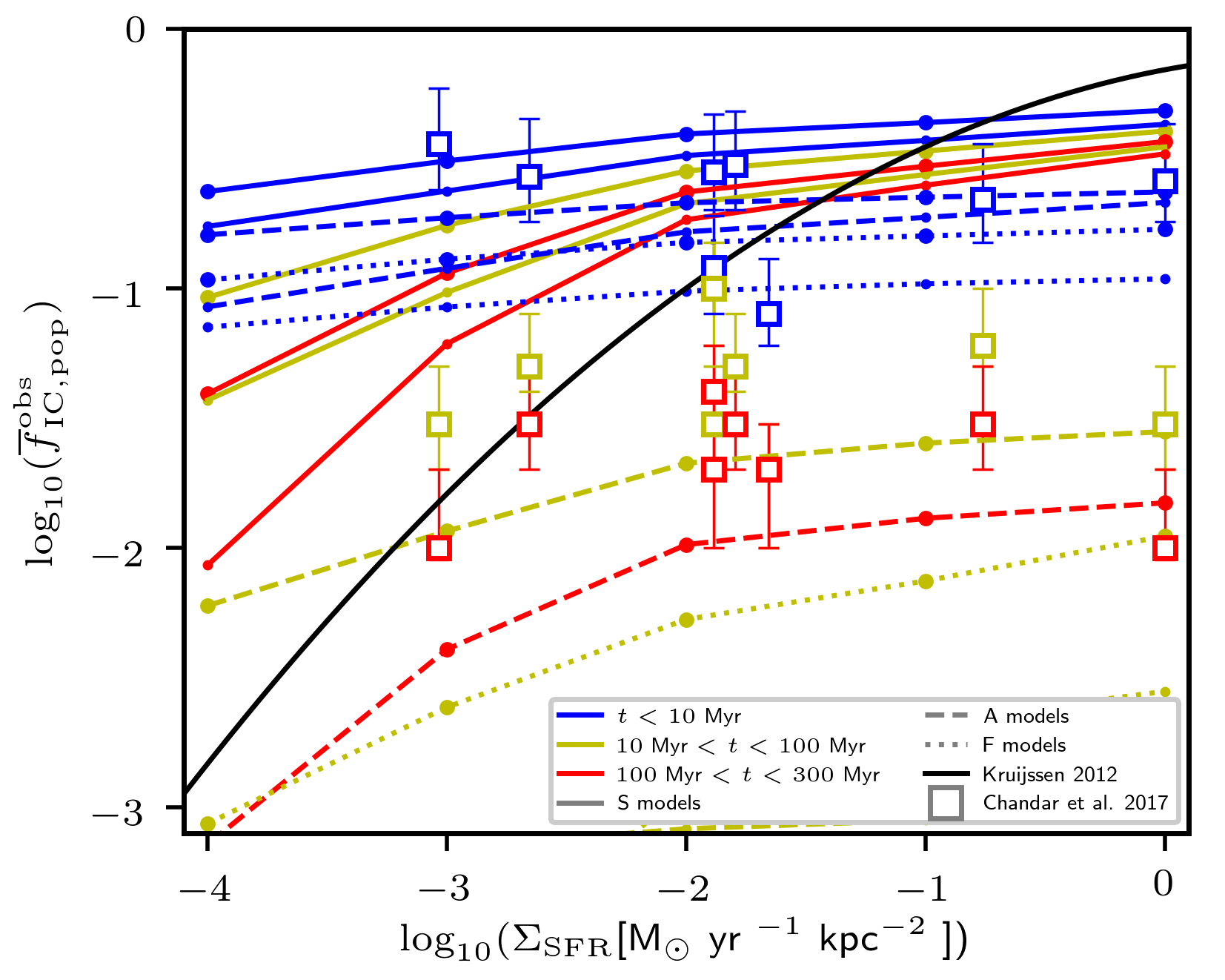}
\caption{
The fraction of stars which is likely to be observed in clusters of age $t < 10 \Myr$ (blue lines), $10 \Myr < t < 100 \Myr$ (yellow lines),
and $100 \Myr < t < 300 \Myr$ (red lines) as a function of $\Sigma_{\rm SFR}$.
The plot shows results for "S", "A" and "F" models for the ECMF of slope $\beta = 2$ by solid, dashed and dotted lines, respectively. 
The lines are shown for $r_{\rm search} = 2 \Pc$ (smaller dots) and $r_{\rm search} = 5 \Pc$ (larger dots) for each model.
The prediction of the \GamSig model and the observational data due to \citet{Chandar2017} are shown by the black line and by squares, respectively.
}
\label{fsigmaSFR_discuss}
\end{figure} \else \fi

Figure \ref{fsigmaSFR_discuss} shows the fraction of stars which is likely to be observed in clusters for the "S", "A" and "F" models 
assuming the IGIMF theory and $\beta = 2$.
For the youngest clusters ($t \lesssim 10 \Myr$; blue lines), the difference between the models varies less than by a factor of three; 
it decreases from $\overline{f}_{\rm IC,pop}^{\rm obs} (t < 10 \Myr) \approx 0.35$ 
for S models to $\overline{f}_{\rm IC,pop}^{\rm obs} (t < 10 \Myr) \approx 0.13$ for F models (for Milky Way-like $\Sigma_{\rm SFR}$ of $\approx 1 \Mspy$).
Similar to the "S" models, both "A" and "F" models reasonably agree with the observations, 
and they show a much weaker dependence on $\Sigma_{\rm SFR}$ than predicted by the \GamSig theory. 

For older clusters (age $\gtrsim 10 \Myr$; yellow and red lines in \reff{fsigmaSFR_discuss}), "A" and "F" models result 
in significantly lower observed fractions of stars in clusters than in the standard model. 
This brings "A" and "F" models closer to the observations of \citet{Chandar2017} also at later times;
the modelled values of $\overline{f}_{\rm IC,pop}^{\rm obs} (10 \Myr < t < 100 \Myr)$ 
and $\overline{f}_{\rm IC,pop}^{\rm obs} (t < 300 \Myr)$ are even lower than the observations. 
"F" models have a too low fraction of stars in clusters. 
This indicates that the observational data could be well reproduced also for clusters older than $10 \Myr$ and as old as at least $300 \Myr$ 
by either steepening the relationship between the cluster mass and its radius, or by lowering the $\sfe$ (albeit the latter possibility 
needs the $\sfe$ somewhere between $0.25$ and $0.33$) without considering any additional cluster dissolving mechanism such as interactions with molecular clouds. 
This is a remarkable property of "A" and "F" models because standard models necessitate 
an additional cluster dissolving mechanism to explain the observations for $t \gtrsim 10 \Myr$ (\refs{ssCombObs}).
Given the large sensitivity of the fraction of stars in clusters on the $\sfe$ and possibly $\tau_{\rm M}$, it is probable that the observations for clusters 
up to the age of at least $300 \Myr$ 
can be explained by more combinations of the cluster parameters ($a_{\rm ecl}$, $\sfe$ and $\tau_{\rm M}$) than considered in this work.

The impact of gas expulsion on a cluster is a strong function of the ratio $\tau_{\rm M}/t_{\rm cross}$ \citep{Lada1984,Baumgardt2007}. 
For all the clusters studied in this work, gas expulsion is impulsive ($\tau_{\rm M}/t_{\rm cross} \lesssim 1$). 
Assuming $\tau_{\rm M} = 10 \Kms$ independent of $M_{\rm ecl}$, gas expulsion becomes adiabatic ($\tau_{\rm M}/t_{\rm cross} \gg 1$) 
if $M_{\rm ecl} \gg 10^4 \Msun$. 
Adiabatic gas expulsion unbinds a substantially lower fraction of stars than impulsive gas expulsion \citep[e.g.][]{Lada1984}.
Since we extrapolate results from clusters with impulsive gas expulsion towards the more massive ones with adiabatic gas expulsion, 
$f_{\rm IC} (M_{\rm ecl})$ is underestimated for $M_{\rm ecl} \gtrsim 10^4 \Msun$. 
We estimate that this effect is of secondary importance (at least for the clusters in the youngest age bin) 
because $f_{\rm IC} (M_{\rm ecl} = 6400 \Msun) \gtrsim 0.8$ (\reff{fIndividualClStars}). 
Even if $f_{\rm IC} (M_{\rm ecl})$ were $1$ for $M_{\rm ecl} \gtrsim 10^4 \Msun$, $\overline{f}_{\rm IC,pop} (t < 10 \Myr)$ cannot differ by more than 
a factor of $1/0.8 = 1.25$ from the present models.

The models with $\sfe = 0.25$ present a lower estimate on the $\sfe$ because they underestimate the fractions of stars in clusters after 
$10 \Myr$ (\reff{fsigmaSFR_discuss}). 
What is the expected fraction of stars in clusters in the other extreme if we consider an absence of gas expulsion (i.e. $\sfe = 1$)? 
In this case, the clusters would be unaffected, so they would resemble their state at $t = 0$, at which time $\overline{f}_{\rm IC,pop}^{\rm obs} (0) = 0.61$ 
for a Milky Way-like galaxy ($\Sigma_{\rm g} = 1 \Sd$ and $\beta = 2$).
The present simulations predict $\overline{f}_{\rm IC,pop}^{\rm obs} (t < 10 \Myr) \approx 0.35$.
Thus, the uncertainty in gas expulsion parameters can increase the value of $\overline{f}_{\rm IC,pop}^{\rm obs} (t < 10 \Myr)$ by at most a factor of $1.8$.

\iffigscl
\begin{figure}
\includegraphics[width=\columnwidth]{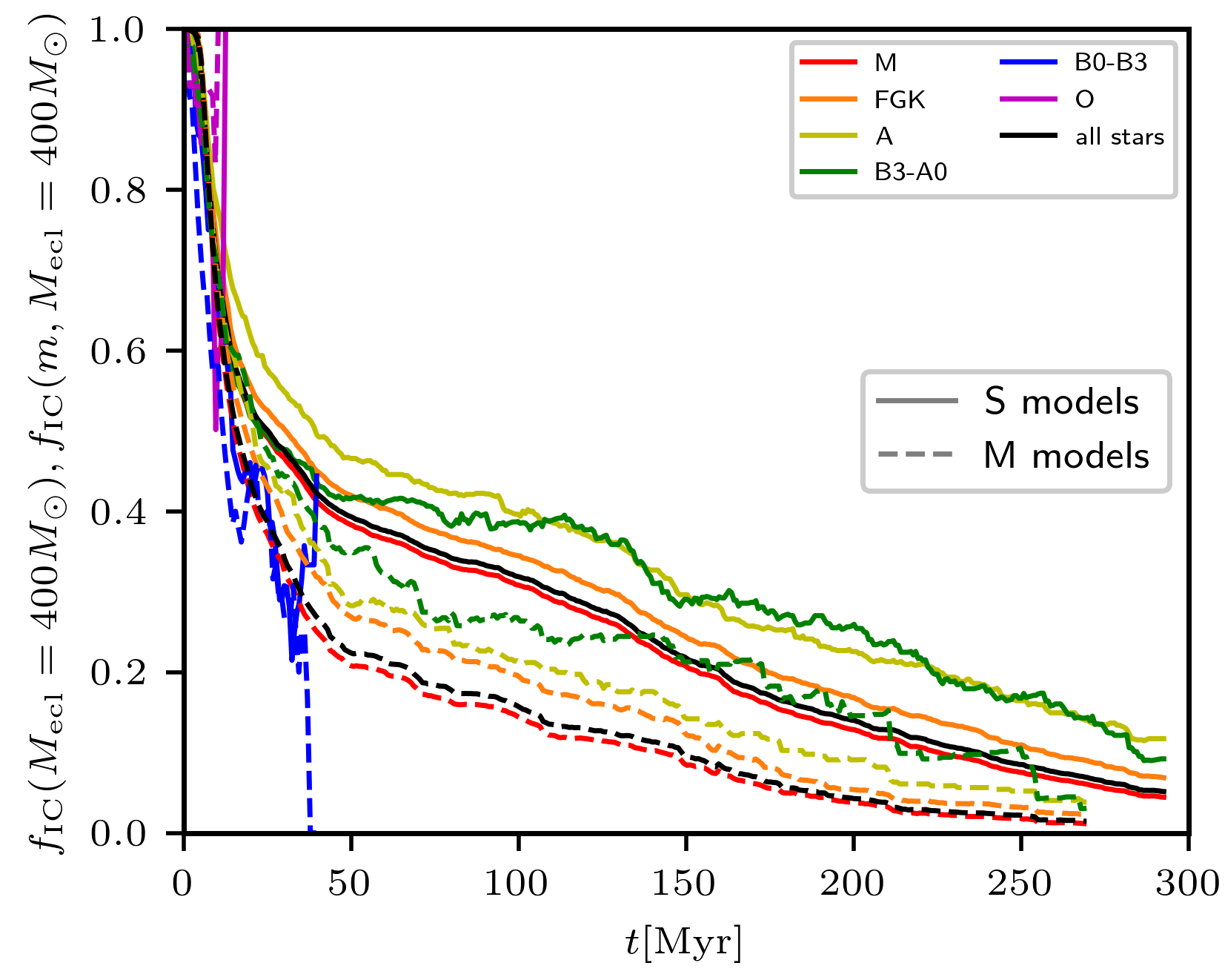}
\caption{The time dependence of $f_{\rm IC}(m)$ for clusters of mass $400 \Msun$ starting with initial conditions for 
models "S" (solid lines), and "M" (dashed lines).
The values of $f_{\rm IC}(m)$ for different stellar spectral types is shown by line colours. 
The fraction of all stars in clusters $f_{\rm IC}(M_{\rm ecl} = 400 \Msun)$ is shown by the black lines.
The clusters orbit the galaxy at $R_{\rm g} = 8 \Kpc$, and they have metallicity $Z = 0.014$.
}
\label{fInfluenceICs}
\end{figure} \else \fi


\subsubsection{Primordial mass segregation}

To study the possible influence of the initial conditions on the fraction of stars to be found in clusters, we perform additional models 
with a high degree of primordial mass segregation (models "M"), which are detailed in \refs{sssAdditionalModels}. 
These models present likely the extreme cases of stellar concentrations within the clusters, having 
a very compact core of massive stars, while the standard ("S" models) are of a more uniform density. 
Figure \ref{fInfluenceICs} compares the time dependence of $f_{\rm IC}(m)$ for a cluster of mass $400 \Msun$ 
between "S" and "M" models. 

Generally, all the models have $f_{\rm IC}(m)$ increasing with $m$ at a given time, which reflects mass segregation, which establishes dynamically in models "S". 
The primordially mass segregated clusters "M" lose stars more quickly and dissolve earlier than their non-mass segregated counterparts "S". 
Assuming that we observe clusters of mass $400 \Msun$, which formed in a galaxy with a constant SFR over a long time interval (longer than $300 \Myr$), 
the fraction of stars found in these clusters is $\overline{f}_{\rm IC}(t < 300 \Myr) = 0.26$ for "S" 
models and $\overline{f}_{\rm IC}(t < 300 \Myr) = 0.17$ for "M" models. 
These values are for all stars in gravitationally bound objects, many of which are comprised from only tens of stars, and are not likely to be observed.
The same estimate of the fraction of observable stars in clusters as done in \refs{ssCombObs} yields 
for "S" models $\overline{f}_{\rm IC}^{\rm obs} (t < 10 \Myr) = 0.34$ 
and $\overline{f}_{\rm IC}^{\rm obs} (t < 10 \Myr) = 0.49$ for $r_{\rm search} = 2 \Pc$ and $5 \Pc$, respectively. 
"M" models have slightly lower fractions of the youngest stars in clusters, $\overline{f}_{\rm IC}^{\rm obs} (t < 10 \Myr) = 0.31$ 
and $\overline{f}_{\rm IC}^{\rm obs} (t < 10 \Myr) = 0.48$ for $r_{\rm search} = 2 \Pc$ and $5 \Pc$, respectively.
Thus, primordial mass segregation decreases $\overline{f}_{\rm IC}$ in a population of clusters, but its observed manifestation in the youngest age group is modest.

We expect that the difference of $\overline{f}_{\rm IC}^{\rm obs} (t < 10 \Myr)$ between "M" and "S" models decreases for clusters more massive than $400 \Msun$ because 
more massive clusters tend to retain more of their stars, and the difference increases 
for clusters of mass below $400 \Msun$ for the opposite reason
\footnote{We choose the cluster mass to be $400 \Msun$ because it lies approximately in between the interval $(\log_{10}(M_{\rm ecl,min}), \log_{10}(M_{\rm ecl,max}))$, 
i.e. the total number of stars contained in clusters forming within mass interval $(M_{\rm ecl,min}, 400 \Msun)$ and $(400 \Msun, M_{\rm ecl,max})$ 
is comparable (assuming $\beta \approx 2$).}
. 
In total, we expect that for a population of primordially mass segregated clusters 
with $\beta = 2$, $\overline{f}_{\rm IC}^{\rm obs} (t < 10 \Myr)$ is slightly lower (approximately up to a factor of $1.2$) than 
for the non-mass segregated clusters.

\subsection{The influence of the form of the ECMF}

The admitted variation of the slope $\beta$, which ranges from $1.8$ to $2.2$ in most galaxies \citep[e.g.][]{Whitmore1999,Lada2003,Chandar2010}, 
results in the 'observed' variation of the fraction of stars in the youngest clusters to vary in the range of 
$\overline{f}_{\rm IC}^{\rm obs} (t < 10 \Myr) = 0.18$ to $0.55$ (for a Milky Way-like galaxy with $\Sigma_{\rm g} = 1 \Sd$).
It differs by a factor of $2$ from the standard value for $\beta = 2$. 
This uncertainty is of a comparable magnitude as the uncertainty in the initial cluster conditions. 

Another important parameter of the ECMF is the cluster lower mass limit $M_{\rm ecl,min}$. 
\citet{Chandar2017} use $M_{\rm ecl,min} = 100 \Msun$, while the formation of clusters of mass significantly lower than that is known 
\citep[e.g.][]{Kroupa2003a,Kirk2011,Joncour2018,Plunkett2018}. 
Decreasing the lower mass cluster limit results in a higher value of $\Gamma$. 
For example, for the LMC \citet{Chandar2017} obtain $\Gamma = 0.27$, while it would be $\Gamma = 0.41$ for $M_{\rm ecl,min} = 5 \Msun$.
Taking into account a lower mass limit $M_{\rm ecl,min}$ would bring their observations to a better agreement with our models.


\subsection{Comparison to observations}

An important step in determining the value of $\Gamma$ is to identify all clusters above a given mass limit $M_{\rm ecl,lim}$ 
(typically $M_{\rm ecl,lim} > M_{\rm ecl,min}$; \citealt{Goddard2010,Chandar2017}). 
Then, the total cluster population is obtained by extrapolating 
the number of clusters towards $M_{\rm ecl}$ lower than $M_{\rm lim}$ assuming an ECMF slope $\beta$. 
Even clusters with mass above $M_{\rm ecl,lim}$ suffer from incompleteness mainly due to extinction, where some clusters are obscured by molecular clouds, 
and crowding in the most active star forming regions. 
These observational limitations tend to underestimate $\Gamma$. 
Taking this into account could improve the agreement between our model and observations in galaxies with higher SFRs as seen in \reff{fsigmaSFR}.

\section{Conclusions}

\label{sSummary}

We have investigated whether the majority of stars form in gravitationally bound or unbound systems using N-body simulations. 
As a starting hypothesis, we assumed that $\Gamma = 1$, that is, that all stars form in gravitationally bound embedded clusters. 
This hypothesis is consistent with and motivated by  
observational data of young star forming regions. 
The clusters experience a gravitational potential drop as the result of early gas expulsion, 
which impacts low-mass clusters more strongly than higher-mass clusters.
Additional stars escape their initial host clusters due to ejection, evaporation and the tidal field of the host galaxy. 

The time evolution of the fraction of stars in clusters for all clusters originating in a single coeval star burst is shown in \reff{fpopulationIC_betaSFR}. 
For example, in a Milky Way-like galaxy ($\sfr \approx 1 \Mspy$) with the ECMF slope $\beta = 2$, our simulations predict that $35$ \% 
of stars remain gravitationally bound to their initial host clusters at the age of $200 \Myr$. 
More massive stars are more likely to be present in clusters (83\% of O-type stars) than lower-mass stars (32\% of M-type stars younger than $300 \Myr$; 
\reft{tbetaSFR}). 

The galactic tidal field disrupts clusters more easily at smaller galactocentric distances $R_{\rm g}$; assuming the same upper cluster mass 
limit throughout the galaxy, the fraction of stars in clusters increases by a factor of $1.6$ between $R_{\rm g} = 4 \Kpc$ 
and $R_{\rm g} = 12 \Kpc$ for $\beta = 2$ (\reft{tbetaSFR}).
However, in a more realistic scenario where the upper cluster mass limit decreases with increasing $R_{\rm g}$, 
the fraction of stars in clusters decreases by a factor of $1.2$ from $R_{\rm g} = 4 \Kpc$ to $R_{\rm g} = 12 \Kpc$, 
therefore more than compensating for the impact of the tidal field.
The influence of cluster metallicity is negligible. 
If clusters form primordially mass segregated, the fraction of stars in clusters is slightly lower, but not likely more than by a factor of $1.2$. 

We compare our simulations to observations. 
This comparison is deemed approximate because of the systematic differences 
in detecting star clusters and their member stars observationally vs. within our simulations.
For the youngest stars ($t < 10 \Myr$), our models are in agreement with most of the observational data of \citet{Chandar2017} (\reff{fsigmaSFR}). 
Thus, the observations are in agreement with the hypothesis that all stars form in gravitationally bound embedded clusters. 
We further find a rather weak dependence of the fraction $\overline{f}_{\rm IC, pop}^{\rm obs}(t < 10 \Myr)$ of the observed youngest 
stars in clusters on $\Sigma_{\rm SFR}$, which contrasts with the strong dependence posited by the \GamSig model \citep{Kruijssen2012}. 
These results show only weak sensitivity on the particular choice of parameters for our model such as the initial cluster radius or the SFE.
The physical reason for the dependence of $\overline{f}_{\rm IC, pop}^{\rm obs}$ on $\Sigma_{\rm SFR}$ is different from the dependence in the \GamSig model: in our model, it is due to
the lesser impact of gas expulsion on more massive clusters, which form predominantly in galaxies with higher $\Sigma_{\rm SFR}$, 
whereas it is due to star formation at higher gas densities, higher SFEs, 
and higher fraction of stars forming in bound clusters in the \GamSig model.

At later times (ages between 10 to 100 $\Myr$ and 100 to 300 $\Myr$; \reff{fsigmaSFR}), 
our standard models have too large $\overline{f}_{\rm IC, pop}^{\rm obs}$ compared to observations.
A similar result was previously found by \citet{Lamers2005}, who did not include early gas expulsion. 
Thus, early gas expulsion does not reconcile this discrepancy, and other mechanisms (e.g. encounters with giant molecular 
clouds) are needed to reduce the number of clusters and bring them closer to the observed values. 
A case in point are the nearby Hyades star cluster, which may be in the process of being disrupted \citep{Jerabkova2021}.

We propose an alternative explanation to this picture. 
The difference between N-body models and observations could be reconciled by adjusting some parameters of our models (\refs{sssInfluenceRadSFE}), 
the most attractive of which is the SFE and the relationship between the cluster half-mass radius and cluster mass.
In this case gas expulsion, cluster dynamics and galactic potential alone bring the 
results close to observations (at least for clusters younger than $300 \Myr$) without the necessity of employing additional cluster dissolution mechanisms. 

We contrast the dynamical paradigm where all stars form in embedded clusters with the theoretical \GamSig framework that derives the functional form of $\Gamma(\Sigma_{\rm SFR})$ 
by combining the observed and modelled properties of the ISM and star forming regions.
We identify several questionable points in the \GamSig theory, particularly the substitution of several equations into one another, 
some of which are non-linear and/or contain uncertain or poorly constrained terms.
Additionally, the \GamSig framework neglects early stellar feedback, Larson's relations, the time-dependent star formation rate 
during cluster formation, stellar dynamics, and the density threshold for star formation. 
Using a toy model, we show (in \refs{ssGamSigLimitations}) that even small changes (consistent with observations) to the equations assumed 
by the \GamSig model result in very different functional forms of $\Gamma$ as a function of $\Sigma_{\rm SFR}$. 
This leads us to challenge the key prediction of the \GamSig model: that 
most stars form as not gravitationally bound to embedded star clusters in Milky Way-like galaxies. 
Instead, we suggest that cluster dynamics, early gas expulsion, 
and the influence of realistic galactic environments (potential and giant molecular clouds) provide an explanation for the fraction of 
stars observed in star clusters under the hypothesis that all stars originate from gravitationally bound systems.

\begin{acknowledgements}
We would like to thank an anonymous referee for useful comments, which significantly improved the paper. 
We thank Sverre Aarseth for continuously developing the code \nbdvi as well as for his numerous pieces of advice. 
FD acknowledges the European Southern Observatory in Garching where part of this work was completed under the Scientific Visitors Programme.
PK acknowledges support from the Grant Agency of the Czech Republic under grant number 20-21855S.
RIA acknowledges funding provided by an SNSF Eccellenza Professorial Fellowship (Grant No. PCEFP2\_194638) 
and funding from the European Research Council (ERC) under the European Union's Horizon 2020 research and innovation programme (Grant Agreement No. 947660). 
We appreciate the support of the ESO IT team, which was vital for performing the presented numerical models. 
This research made use of Matplotlib Python Package \citep{matplotlib2007}.
\end{acknowledgements}

%
%




\bibliographystyle{aa} 
\bibliography{starsInClusters} 


\begin{appendix}

\section{Estimates of some observed quantities}

Figures \ref{fIndividualClStars} and \ref{fpopulationIC_betaSFR} show the time dependence of the physical fraction of stars located within 
gravitationally bound clusters. 
However, many of these clusters contain only a small number of late type stars, and are not likely to be observed.
The fraction of stars to be observed according to the criteria in \refs{ssCombObs} is plotted in \reff{fobserved}. 

\iffigscl
\begin{figure*}
\includegraphics[width=\textwidth]{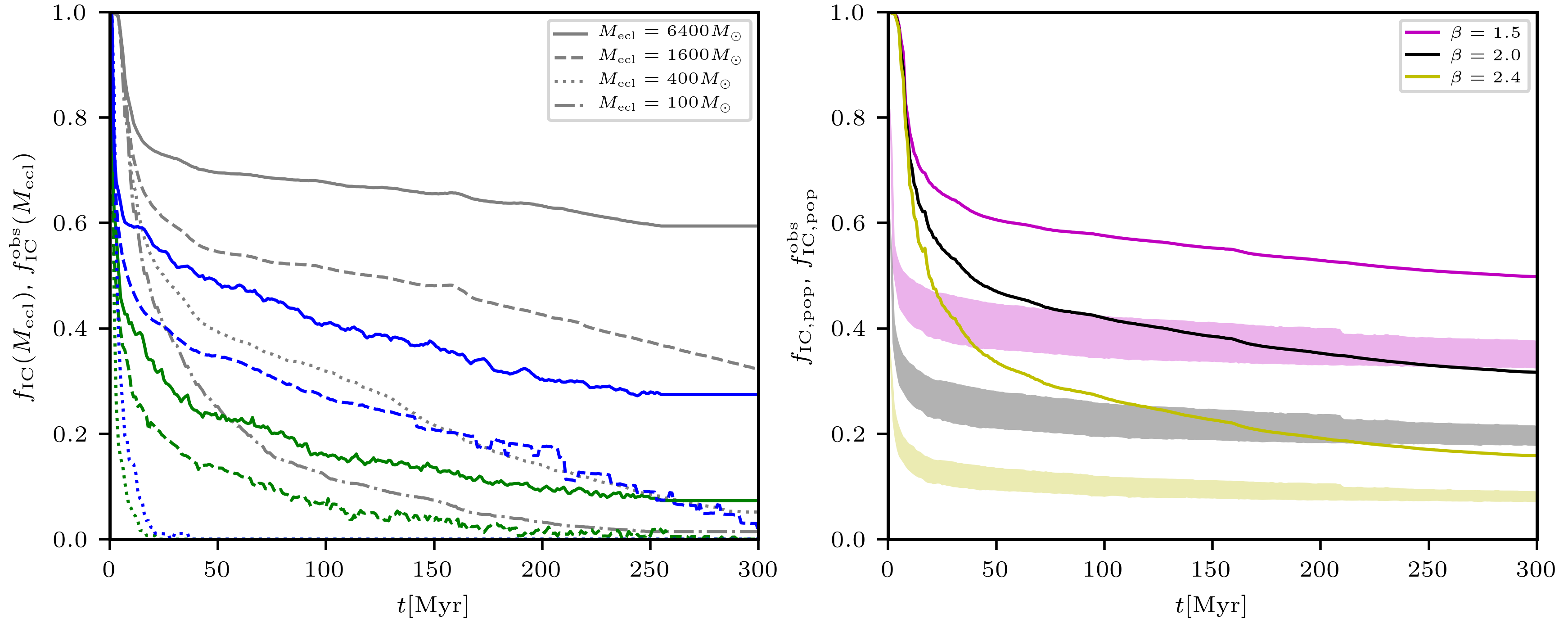}
\caption{\figpan{Left panel:} The physical fraction of stars in clusters (gray lines; the same as \reff{fIndividualClStars}), 
and the estimated observed fraction of stars in clusters 
when using $r_{\rm search} = 5 \Pc$ (blue lines) and $r_{\rm search} = 2 \Pc$ (green lines). 
The cluster mass is indicated by the line style.
\figpan{Right panel:} The physical fraction of stars in clusters (solid lines), 
and the estimated observed fraction of stars in clusters assuming $r_{\rm search}$ between $2$ to $5 \Pc$ (filled areas). 
This panel is for the ECMF of \eq{eicmf} with $\beta = 1.5$ (magenta), $\beta = 2$ (black), and $\beta = 2.4$ (yellow).
}
\label{fobserved}
\end{figure*} \else \fi

\end{appendix}

\end{document}